\documentclass[a4paper,fleqn,usenatbib]{mnras}

\usepackage{newtxtext,newtxmath}

\usepackage[T1]{fontenc}
\usepackage{ae,aecompl}

\usepackage{graphicx}	
\usepackage{amsmath}	
\usepackage{amssymb}	

\newcommand{\se}[1]{\S\ref{sec:#1}}

\newcommand{\Fig}[1]{Figure~\ref{fig:#1}}

\newcommand{\tab}[1]{Table~\ref{tab:#1}}
\newcommand{\be}{\begin{equation}}
\newcommand{\ee}{\end{equation}}
\newcommand{\bea}{\begin{eqnarray}}
\newcommand{\eea}{\end{eqnarray}}

\newcommand{\msun}{{\rm M}_\odot}

\newcommand{\ifm}[1]{\relax\ifmmode#1\else$\mathsurround=0pt #1$\fi}
\newcommand{\kms}{\ifmmode\,{\rm km}\,{\rm s}^{-1}\else km$\,$s$^{-1}$\fi}
\newcommand{\hmpc}{\,\ifm{h^{-1}}{\rm Mpc}}

\newcommand{\Mpc}{\,{\rm Mpc}}

\newcommand{\am}{\mbox{\normalfont\AA}}

\newcommand{\ltsima}{$\; \buildrel < \over \sim \;$}
\newcommand{\lsim}{\lower.5ex\hbox{\ltsima}}
\newcommand{\gtsima}{$\; \buildrel > \over \sim \;$}
\newcommand{\gsim}{\lower.5ex\hbox{\gtsima}}

\def\la{\langle}

\def\omm{\Omega_{\rm m}}

\def\omb{\Omega_{\rm b}}

\def\Ms{M_*}

\usepackage{color}

\title[FL III]{FirstLight III: Rest-frame UV-optical spectral energy distributions of simulated galaxies at cosmic dawn}

\author[Ceverino et al.]{
Daniel Ceverino,$^{1,2,3}$\thanks{E-mail: ceverino@uni-heidelberg.de}
Ralf S. Klessen,$^{1,4}$ Simon C. O. Glover,$^{1}$
\\
$^{1}$Universit\"{a}t Heidelberg, Zentrum f\"{u}r Astronomie, Institut f\"{u}r Theoretische Astrophysik, Albert-Ueberle-Str. 2, 69120 Heidelberg, Germany\\
$^{2}$Cosmic Dawn Center (DAWN) \\
$^{3}$Niels Bohr Institute, University of Copenhagen, Lyngbyvej 2, 2100, Copenhagen $\mbox{\normalfont\O}$, Denmark \\
$^{4}$Universit\"{a}t Heidelberg, Interdisziplin\"{a}res Zentrum f\"{u}r Wissenschaftliches Rechnen, INF 205, 69120, Heidelberg, Germany}

\date{Accepted XXX. Received YYY; in original form ZZZ}

\pubyear{2018}

\begin{document}
\label{firstpage}
\pagerange{\pageref{firstpage}--\pageref{lastpage}}
\maketitle

\begin{abstract}
Using the FirstLight database of 300 zoom-in cosmological simulations we provide rest-frame UV-optical spectral energy distributions of galaxies with complex star-formation histories that are coupled to the non-uniform gas accretion history of galactic halos during cosmic dawn.
The population at any redshift is very diverse ranging from starbursts to quiescent galaxies even at a fixed stellar mass.
This drives a redshift-dependent relation between UV luminosity and stellar mass with a large scatter, driven by the specific star formation rate.
The UV slope and the production efficiency of Lyman continuum photons have high values, consistent with dust-corrected observations.
This indicates young stellar populations with low metallicities.
The FirstLight simulations make predictions on the rest-frame UV-optical absolute magnitudes, colors and optical emission lines of galaxies at $z=6-12$ that will be observed for the first time with
JWST and the next generation of telescopes in the coming decade.
\end{abstract}



\section{Introduction}

The period of cosmic dawn, the first billion years in the history of the Universe, is the final frontier for galaxy formation.
Theory roughly  predicts that the first galaxies form in the first gravitationally bound structures at high redshifts, $z \ge 12$.
They quickly start to ionize their surroundings, driving the reionization of the Universe, which ends around $z \ge 6$.
 However, little is known about these primeval galaxies.
 One of their most important properties is their spectral energy distribution (SED).
 It gives information about the stellar populations and the surrounding gas in these primeval galaxies.
 
 Fitting SED templates to rest-frame UV-optical photometry is a common way to estimate the parameters of high-z galaxies, such as stellar mass ($\Ms$) and star formation rate (SFR) \citep{SawickiYee98, Papovich01, Shapley01, Giavalisco02, Stark09, Salmon15}.
 However, this modeling makes assumptions about other properties of the stellar content of these galaxies.
 For example, stellar population models that include the effects of binary evolution or stellar rotation provide a better fit to the UV-optical spectra of high-z galaxies \citep{Steidel16}.
 Simulations of galaxy formation are great tools to test or constrain these assumptions.
They have shown that the inclusion of binary stars yields a harder UV spectrum \citep{Ma16}.
 Therefore, binary evolution has become an important ingredient in the current modeling of the stellar light at cosmic dawn \citep[][and references therein]{Ma18, Rosdahl18}.
 
 One of the key uncertainties in SED modelling is the star formation history (SFH). 
 Due to the lack of constraints, a constant SFR is usually assumed \citep{Salmon15}, but there is no evidence that this is a good assumption.
 Exponentially declining histories usually provide a worse fit for these high-z galaxies \citep{Salmon15, Hashimoto18}.
 On the other hand, many simulations suggest smoothly increasing SFRs with time \citep{Finlator11, Jiang16}, especially during cosmic dawn, when the cosmic SFR density is increasing with time. 
 Recent observations that also fit emission lines can better constrain  the instantaneous SFR and they 
  need more complex SFH with two different populations of different ages \citep{Hashimoto18}.
 Most probably, the SFHs of galaxies at cosmic dawn are not easily parametrized by simple functions, as they reflect the dynamical and violent times of very high gas accretion rates into galaxies.
 
 Cosmological simulations of galaxy formation can provide complex SFHs that are coupled with the non-uniform gas accretion history of galactic halos.
 The FirstLight Simulations \citep[][Paper I]{PaperI} have provided a diversity of non-monotonic SFHs, characterized by bursts of star formation.
 Paper II  \citep{PaperII} found that
 galaxies at redshifts $z=5-15$ spend 70\% of their time in these bursts with very high specific star formation rates (SSFRs).
 The typical burst at $z=6$ has a duration of $\sim$100 Myr, one tenth of the age of the Universe at that time.
 A quarter of the bursts have significantly shorter times scales of 40-80 Myr, mostly driven by mergers.
 Therefore,  it is important to use complex and diverse SFH for the calculation of new SED templates consistent with cosmological mass accretion histories of galactic halos.
 
We aim to understand the physical processes shaping the SEDs of galaxies at cosmic dawn.
As a first step, this paper uses the FirstLight simulations to make predictions of the rest-frame UV-optical SEDs that will be observed for the first time with 
the James Webb Space Telescope (JWST) and
the next generation of telescopes in the coming decade.
The outline of this paper is as follows.
Section \se{runs} presents the FirstLight simulations.
Section \se{SEDmodel} describes the SED modeling.
The results section (\se{R}) includes properties of the UV spectra, such as UV magnitude (\se{MvMs}), slope and  production efficiency of ionizing photons (\se{UV}), the equivalent width of optical emission lines, such as H$\alpha$ (\se{EW}), the absolute magnitude in the V-band (\se{VMs}), color-magnitude and color-color relations (\se{Colors}), and BPT diagrams (\se{BPT}).
Finally, section \se{summary} ends the paper with the summary and discussion.

\section{Simulations}
\label{sec:runs}

This paper uses a complete mass-selected subsample of
galaxies simulated in the FirstLight project described fully in Paper I.
The subsample consists of 290 halos with a maximum circular velocity (V$_{\rm max}$) between 50 and 250 $\kms$, selected at $z=5$. The halos cover a mass range between a few times $10^9$ and a few times $10^{11} \ \msun$. 
This range excludes more massive and rare halos with number densities lower than $\sim 3 \times 10^{-4} (h^{-1} \Mpc)^{-3}$, 
as well as small halos in which galaxy formation is extremely inefficient.
 
The target halos are initially selected using low-resolution N-body only simulations of two cosmological boxes with sizes 10 and 20 $\hmpc$, assuming WMAP5 cosmology with $\omm=0.27$, $\omb=0.045$, $h=0.7$, $\sigma_8=0.82$ \citep{Komatsu09}.  
We select all distinct halos with V$_{\rm max}$ at z = 5 greater than a specified threshold, log $V_{\rm cut}=1.7$ in the 10 $\hmpc$ box and log $V_{\rm cut}=2.0$ in the 20 $\hmpc$ box.
Initial conditions for the selected halos with much higher resolution are then generated using a standard zoom-in technique  \citep{Klypin11}.
 The DM particle mass resolution is m$_{\rm DM}=10^4 \ \msun$. The minimum mass of star particles is $100 \ \msun$.
 The maximum spatial resolution is always between 8.7 and 17 proper pc (a comoving resolution of 109 pc after $z=11$).
  
 The simulations are performed with the  \textsc{ART} code
\citep{Kravtsov97,Kravtsov03, Ceverino09, Ceverino14}, which accurately follows the evolution of a
gravitating $N$-body system and the Eulerian gas dynamics using an adaptive mesh refinement (AMR) approach.
Besides gravity and hydrodynamics, the code incorporates 
many of the astrophysical processes relevant for galaxy formation.  
These processes, representing subgrid 
physics, include gas cooling due to atomic hydrogen and helium, metal and molecular 
hydrogen cooling, photoionization heating by a constant cosmological UV background with partial 
self-shielding, star formation and feedback (thermal+kinetic+radiative), as described in paper I.
The simulations track metals released from SNe-Ia and from SNe-II, using supernovae yields that approximate the  results from \cite{WoosleyWeaver95}. 
These values are given for gas cells and star particles as described in \cite{Kravtsov03}. 

\section{SED modeling}
\label{sec:SEDmodel}

In this paper we focus on the intrinsic SEDs coming from the stellar populations and 
their surrounding, unresolved HII regions.
They provide templates that could be used for a better understanding of the underlying stellar and gas properties in galaxies at cosmic dawn.
Therefore, we ignore dust attenuation or any other radiative transfer effect from intervening gas. 
These effects will be consider in future works. 
Estimations of the dust extinction in other cosmological simulations \citep{Ma18} indicate a significant extinction of $\sim$1 magnitude at 1500 $\am$ for $M_{1500}\simeq-21$.
This corresponds to the brightest galaxies in the FirstLight sample. In most of the snapshots, the effect of dust extinction is much smaller, less than half a magnitude for $M_{1500}\geq-19$, due to the low metallicity in these galaxies.

We compute the SEDs of the simulated galaxies using publicly available tables from the Binary Population and Spectral Synthesis (BPASS) model \citep{Eldridge17} including nebular emission \citep{XiaoStanway18}. 
We only include the effects of binary stars at post-processing, as in other works \citep{Ma18}.
Our model of energy feedback and metal enrichment assumes single-star evolutionary tracks.

\subsection{BPASS}

The stellar spectrum from 1 $\am$ to 100000 $\am$ coming from each star particle uses the templates of single stellar populations (SSP) of BPASS\_v2.1 assuming a Kroupa-like IMF with power slopes $\alpha_1=-1.3$ for star masses $m=0.1-0.5 \ \msun$ and $\alpha_2=-2.35$ for star masses $m=0.5-100 \ \msun$.
We use a grid of 13 values of metallicity, from $Z=10^{-5}$ to 0.04 (Z$_\odot=0.02$),
and 40 logarithmic bins in SSP ages, between 1 Myr and 100 Gyr.

The nebular emission from the same particles
uses 20 bins in SSP ages, from 1 Myr to 100 Myr and the same range of metallicities.
As described in \cite{XiaoStanway18}, the ionization parameter is measured at the Str\"{o}mgren radius and it ranges from $U=-3.5$ to -1.5.  
We assume a constant nebular density of $n_{\rm H}=10^2 \ {\rm cm}^{-3}$, because the simulations do not resolve the nebular regions around young stars where most of the nebular light is emitted. 
Denser HII regions,  $n_{\rm H}=10^3 \ {\rm cm}^{-3}$, give similar results for the SSP metallicities considered in this paper \citep{XiaoStanway18}.
The Str\"{o}mgren radius for each star particle is calculated using the nebular density and the properties of the stellar population.
This publicly available grid of Cloudy models gives the luminosity of the most prominent emission lines in the optical and UV. We also include nebular continuum from hydrogen free-bound and free-free radiation \citep{Osterbrock}.

\subsection{Stellar and nebular metallicities}	

We compute galactic SEDs combining all individual SEDs coming from all star particles within each galaxy. 
We assume different metallicity indicators for stars and nebular regions,
because we cannot assume a solar relative abundance pattern during cosmic dawn.
Observations of stellar and nebular spectra of high-redshift galaxies \citep{Steidel16} suggest that only models that use low iron abundances and relatively high oxygen abundances match all observational constraints. This is motivated from the fact that iron controls the UV stellar opacity and the binary evolution while 
oxygen and other $\alpha$ elements dominate the metal content and cooling of the surrounding nebular gas \citep{Rix04}.
Therefore the stellar metallicity controls the stellar SED and is traced by iron abundances and the nebular metallicity follows the oxygen mass ratio.

We compute the iron abundances from the supernovae yields (SNII and SNIa) assuming: 
1)  iron accounts for half of the metals produced in SNIa explosions \citep{Thielemann86}, 
2) the fraction of iron from the total amount of metals produced in SNII ranges between 2.6\% and 3.3\%, for metal mass ratios between 0 and solar \citep{Nomoto06},
3) the  solar value of the iron mass ratio is $1.23 \times 10^{-3}$ \citep{Grevesse98}.

For nebular metallicities, we assume that the unresolved nebular region around each star particle shares the same mass ratio of metals produced in SNII explosions as in the star particle. 

\subsection{Examples of galactic SEDs at $z$=6}

\begin{figure}
	\includegraphics[width=\columnwidth]{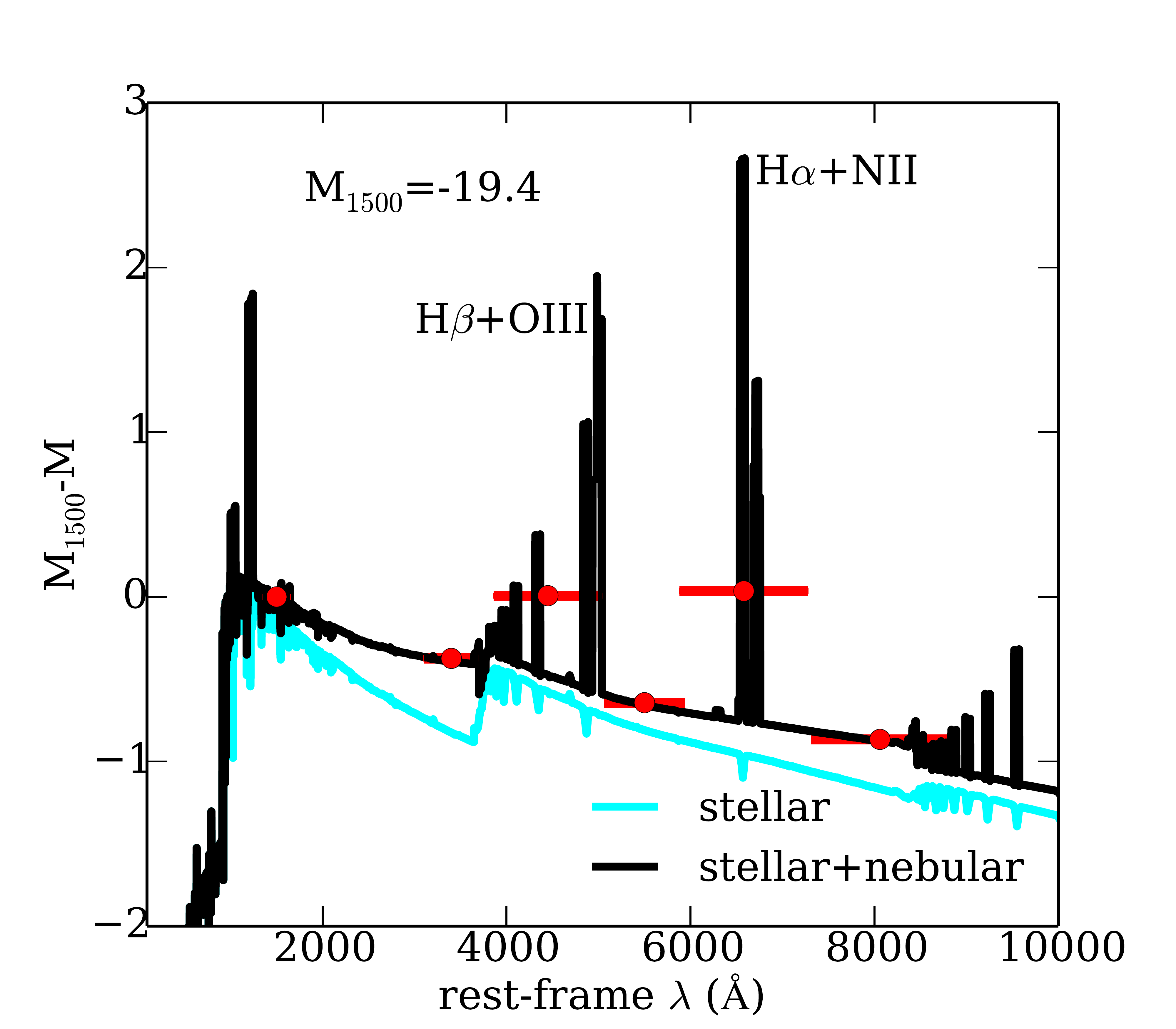}
	\includegraphics[width=\columnwidth]{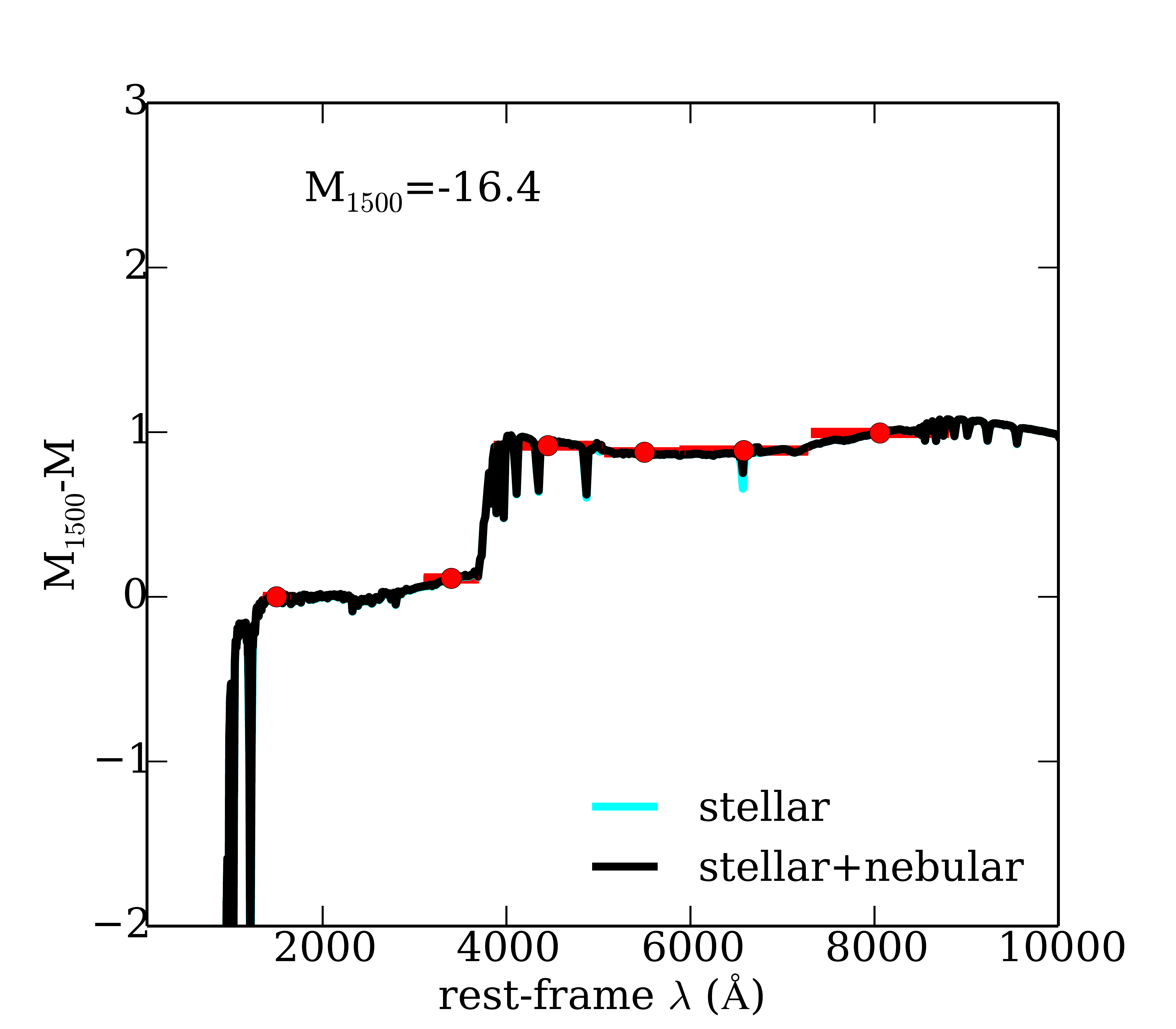}	
	 \caption{SEDs of the same galaxy at the peak of a star formation burst at $z\simeq6$ (top) and 
	 during the subsequent quiescent phase 200 Myr later (bottom). 
	 Cyan lines represent stellar light and black lines include nebular emission (lines+continuum). 
	 Red circles with bars represent the six photometric bands described in \tab{1}.
	 The monochromatic luminosity at each wavelength in erg/s/Hz is normalized to the absolute magnitude at 1500 $\am$ (top label).}
	  \label{fig:SEDs}
\end{figure}

We compute the SEDs for all FirstLight snapshots at every $\sim$10 Myr from $z=15$ to $z=5$,
although the time resolution in the simulations is much shorter, $\sim$1000 yr.
In this paper we focus on the rest-frame UV-optical range from 100 $\am$ to 1 $\mu m$.
The luminosity at each wavelength is normalized to the absolute magnitude at 1500 $\am$, assuming a standard conversion to AB magnitude \citep{Bouwens08}:
\begin{equation} 
M_{1500} - M =  M_{1500}  + 21.91 + 2.5 \ \rm{log} \left( \frac{L_{\nu}}{2.5 \times 10^{29}  \rm {erg} \  \rm{s}^{-1} Hz^{-1}} \right)
\label{eq:MUV}
\end{equation}
We consider six commonly used photometric rest-frame bands (\tab{1}).
They are top-hat filters centered at $\lambda_0$ with a $\Delta \lambda$ bandwidth \citep{Binney}.
They are not chosen to represent a particular set of filters, as we are not comparing with specific observations at this stage.
 They just cover all relevant regions of the rest-frame UV-optical spectrum and they will guide our discussion about the general shapes of the SEDs
 at cosmic dawn.

\begin{table} 
\caption{Photometric bands used in this paper}
 \begin{center} 
 \begin{tabular}{ccc} \hline 
\multicolumn{2}{c} {Band } \ \ $\lambda_0$ (\am) & $\Delta \lambda$ (\am) \\
\hline 
M$_{1500}$ &       	1500  &	  300 	\\
M\_U        &               3400  &      600 	\\
M\_B        &	         4450  &    1180 	\\
M\_V        &               5500  &      880  	\\
M\_R       &               6580  &    1400  	\\
M\_I         &               8060  &    1500          \\
 \end{tabular} 
 \end{center} 
\label{tab:1} 
 \end{table} 

\Fig{SEDs} shows two examples which represent the same galaxy with a stellar mass of $\Ms\simeq10^8 \ \msun$ at $z\simeq6$, at two different phases of evolution.
The top panel depicts the galaxy at the peak of a burst of star formation.  The starburst brings the galaxy a factor 6 times above the star forming main sequence at that redshift (SSFR= 30 Gyr$^{-1}$). Its luminosity at 1500 $\am$ is correspondingly high, $M_{1500}\simeq-19$.
The nebular emission dominates the SED at all wavelengths.
Lines, such as Ly$\alpha$, H$\alpha$, OIII or SiIII, are particularly prominent. 
For example, the equivalent width (EW) of H$\alpha$+NII exceeds 1000 $\am$.
OIII+H$\beta$ is the next prominent feature with ${\rm EW} \simeq700 \am$.
These lines contaminate the continuum in the $B$ and $R$ bands.
The nebular continuum is also particularly important in the $U$ band and it is able to completely remove the mass-sensitive Balmer break at $4000 \am$.
The stellar continuum is also very steep, consistent with a young stellar population with extremely blue colors, $V-I=-0.2$ and $U-V=-0.4$.
These are the typical features of an extreme emission-line galaxy at cosmic dawn.

The SED of the same galaxy 200 Myr after the starburst looks very different (bottom panel of \Fig{SEDs}).
The starburst and the subsequent feedback have quenched star formation  significantly (SSFR= 0.02 Gyr$^{-1}$), placing the galaxy well below the star-forming main sequence (Paper II).
The SED confirms this quiescent nature. 
The luminosity at 1500 $\am$ is very low, $M_{1500}\simeq-16$, for that stellar mass.
There is no  significant nebular emission. Instead, there is a strong Balmer break, typical of a mature population.
The colors are significantly redder,  $V-I=0.2$ and $U-V=0.8$, although the galaxy would not be classified as "red and dead" according to low redshifts color cuts  \citep{Williams09}.
These thresholds may miss a significant fraction of low-mass quenched galaxies \citep{Belli18}.
Therefore, caution is recommended when using low-$z$ classifications for galaxies at cosmic dawn.

\section{Results}
\label{sec:R}

As shown in previous sections, the FirstLight sample covers a large and diverse range of galaxies at cosmic dawn.
It contains both star-forming and quenched galaxies with very different SED properties.

\subsection{Evolution of the $\Ms$-M$_{1500}$ relation}
\label{sec:MvMs}

\begin{figure}
	\includegraphics[width=\columnwidth]{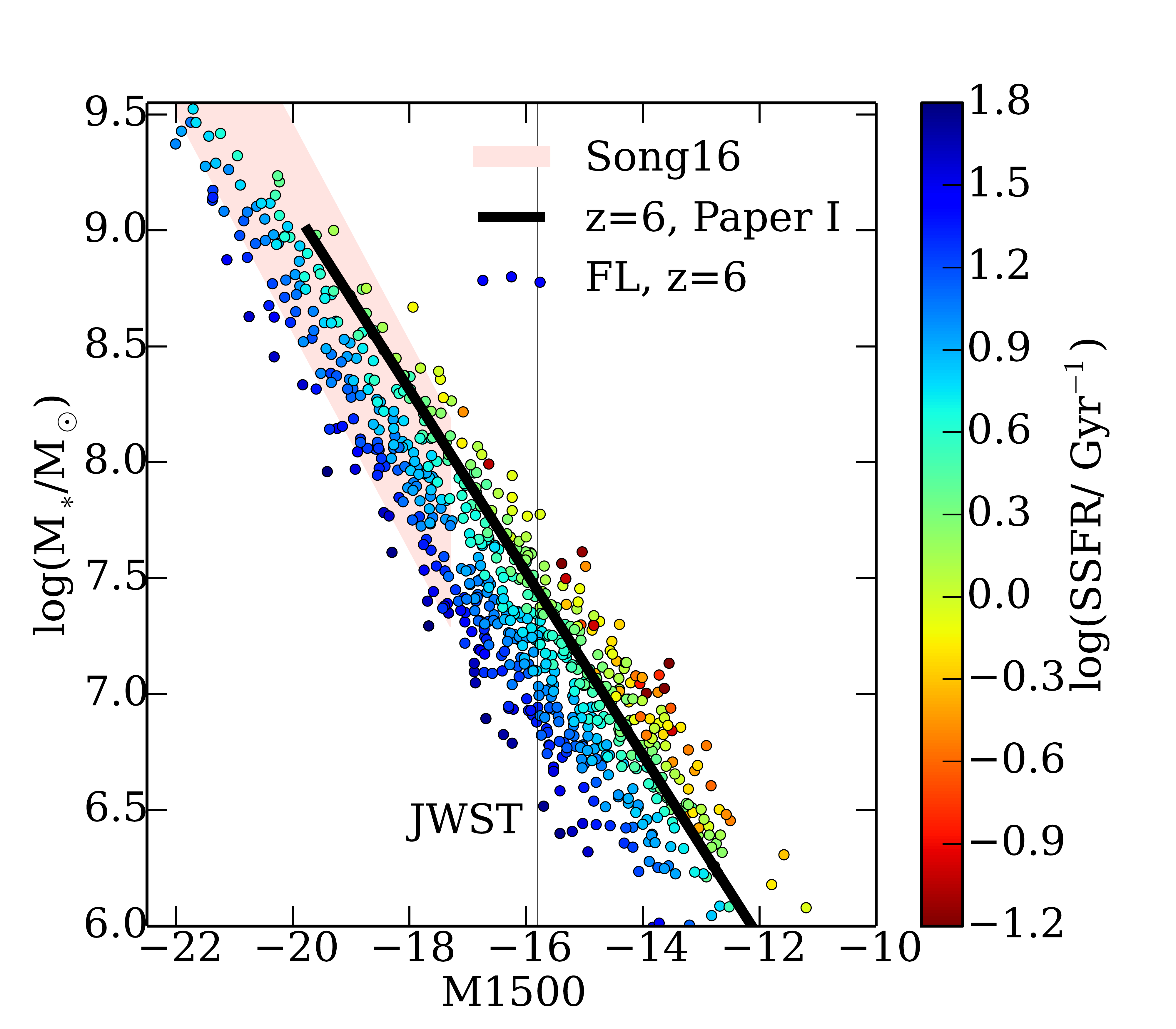}
	\includegraphics[width=\columnwidth]{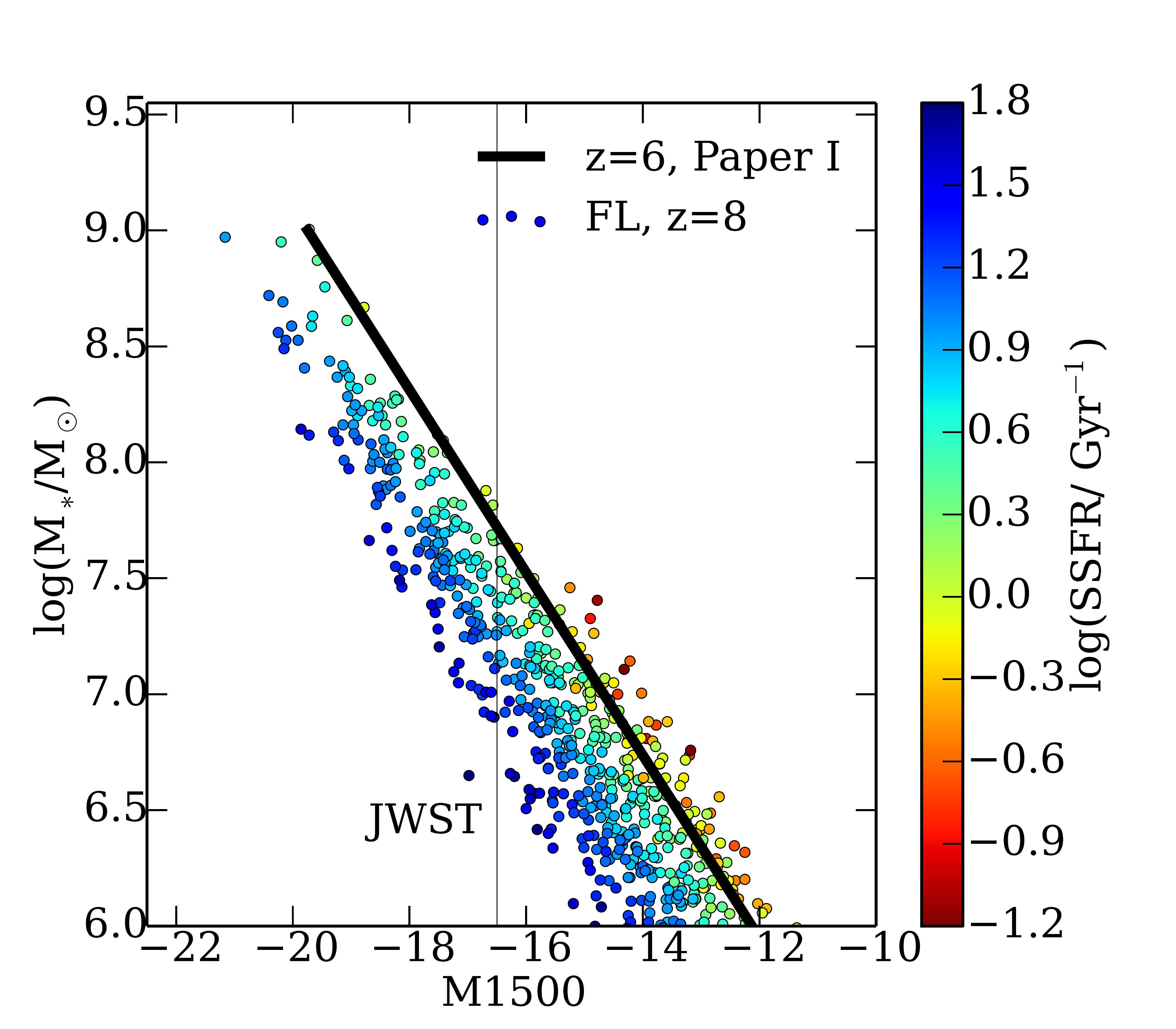}
	\includegraphics[width=\columnwidth]{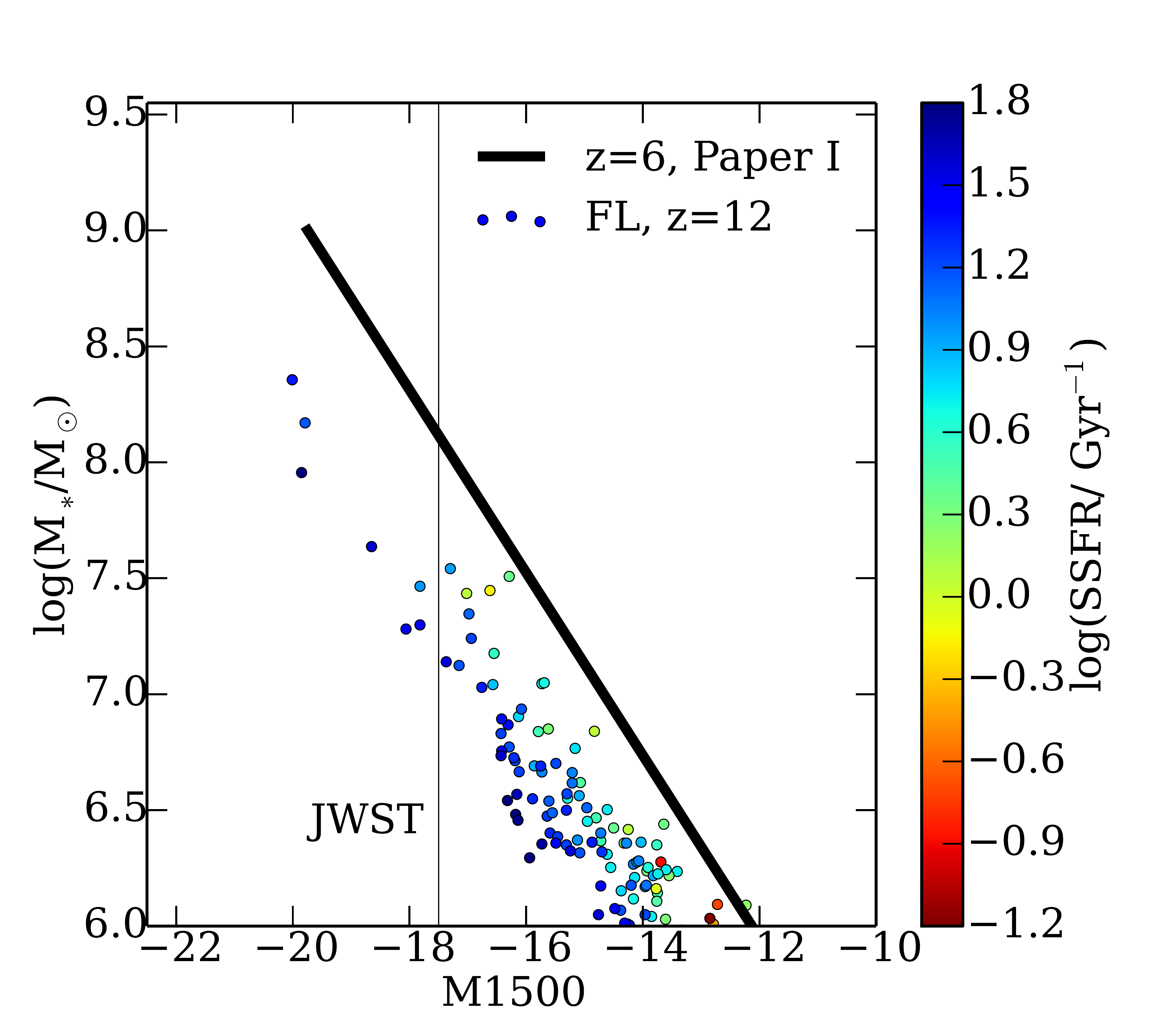}	
	 \caption{Stellar mass versus M$_{1500}$, coloured by SSFR at z=6, 8, and 12.
	 The simulated sample is consistent with current observations \citep{Song16} and with Paper I.
	 Galaxies are gradually brighter at higher z for a fixed mass.
	 A vertical line marks the limit of a ultra-deep JWST field with a limiting magnitude of ${\rm m}_{\rm lim}=31$. }
	  \label{fig:MsMUVSSFR}
\end{figure}

\begin{figure}
	\includegraphics[width=\columnwidth]{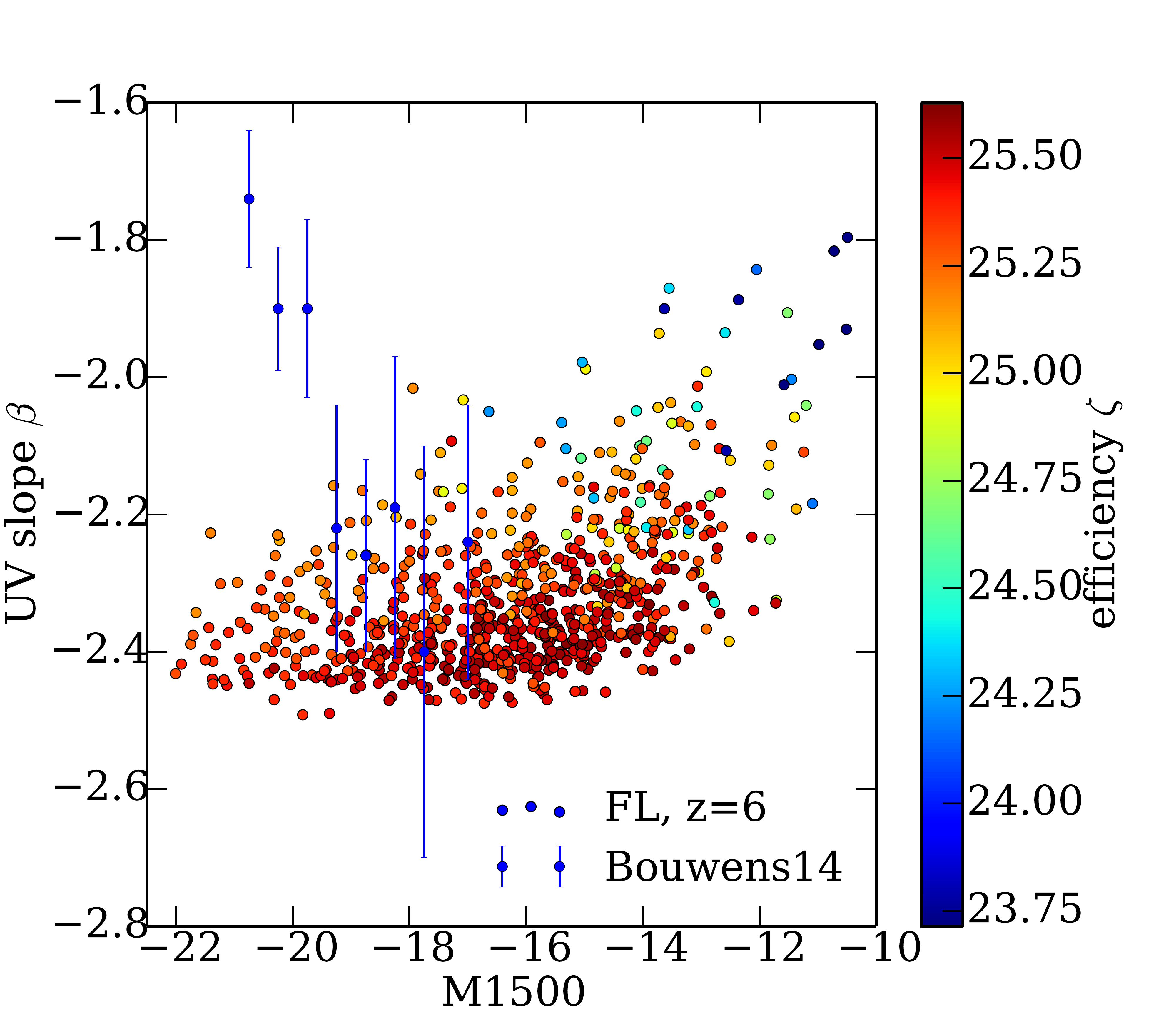}
	\includegraphics[width=\columnwidth]{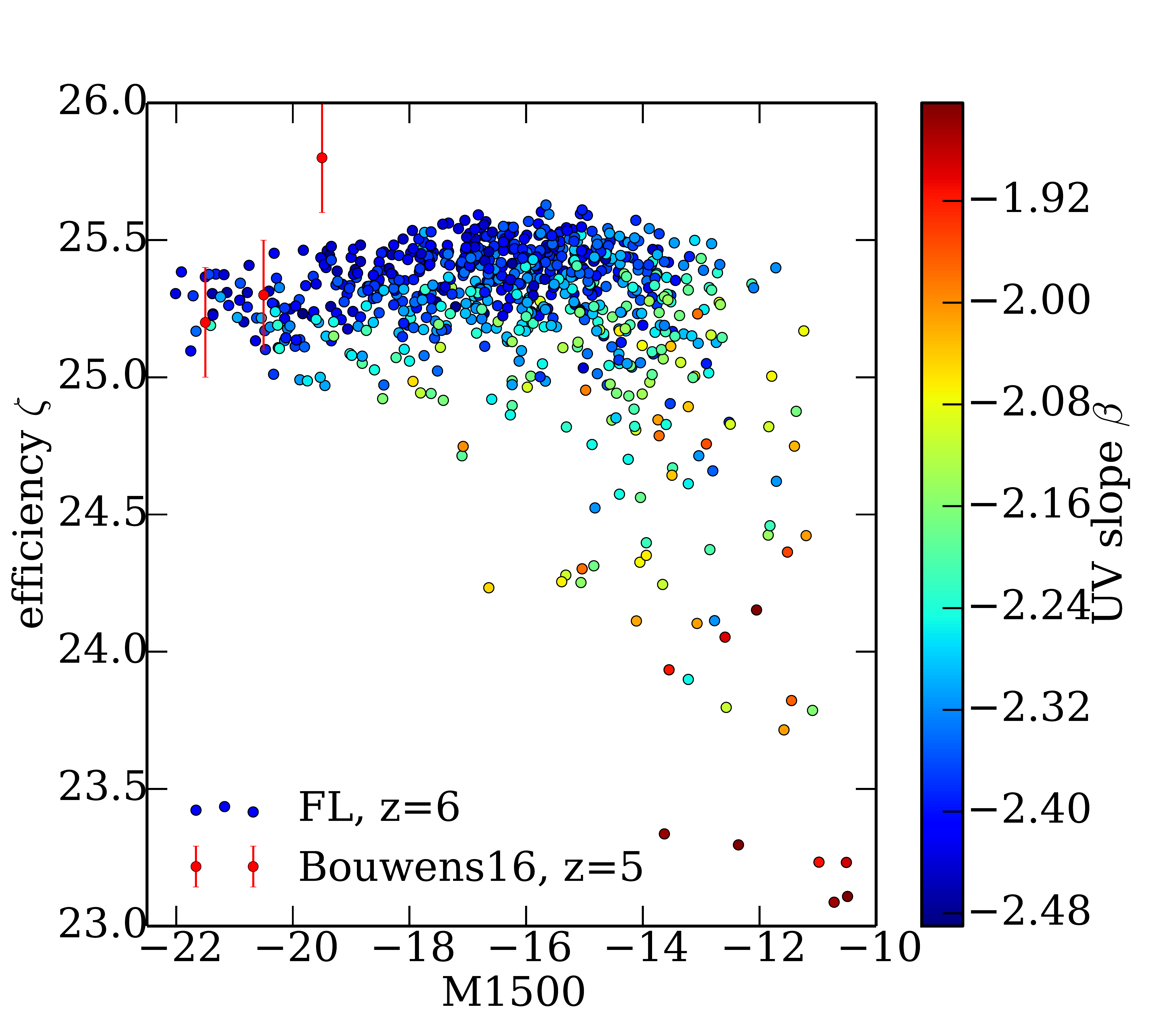}
	 \caption{UV slope ($\beta$) and production efficiency of Lyman continuum photons ($\zeta$) versus $M_{1500}$.
	 Galaxies fainter than -20 have UV slopes consistent with observations \citep{Bouwens14}. 
	 Observations of brighter galaxies need to be corrected for dust attenuation. 
	 $\zeta$ values are consistent with observations \citep{Bouwens16c}, with the exception of the faintest bin. }
	  \label{fig:betazeta}
\end{figure}	

\begin{figure}
	\includegraphics[width=\columnwidth]{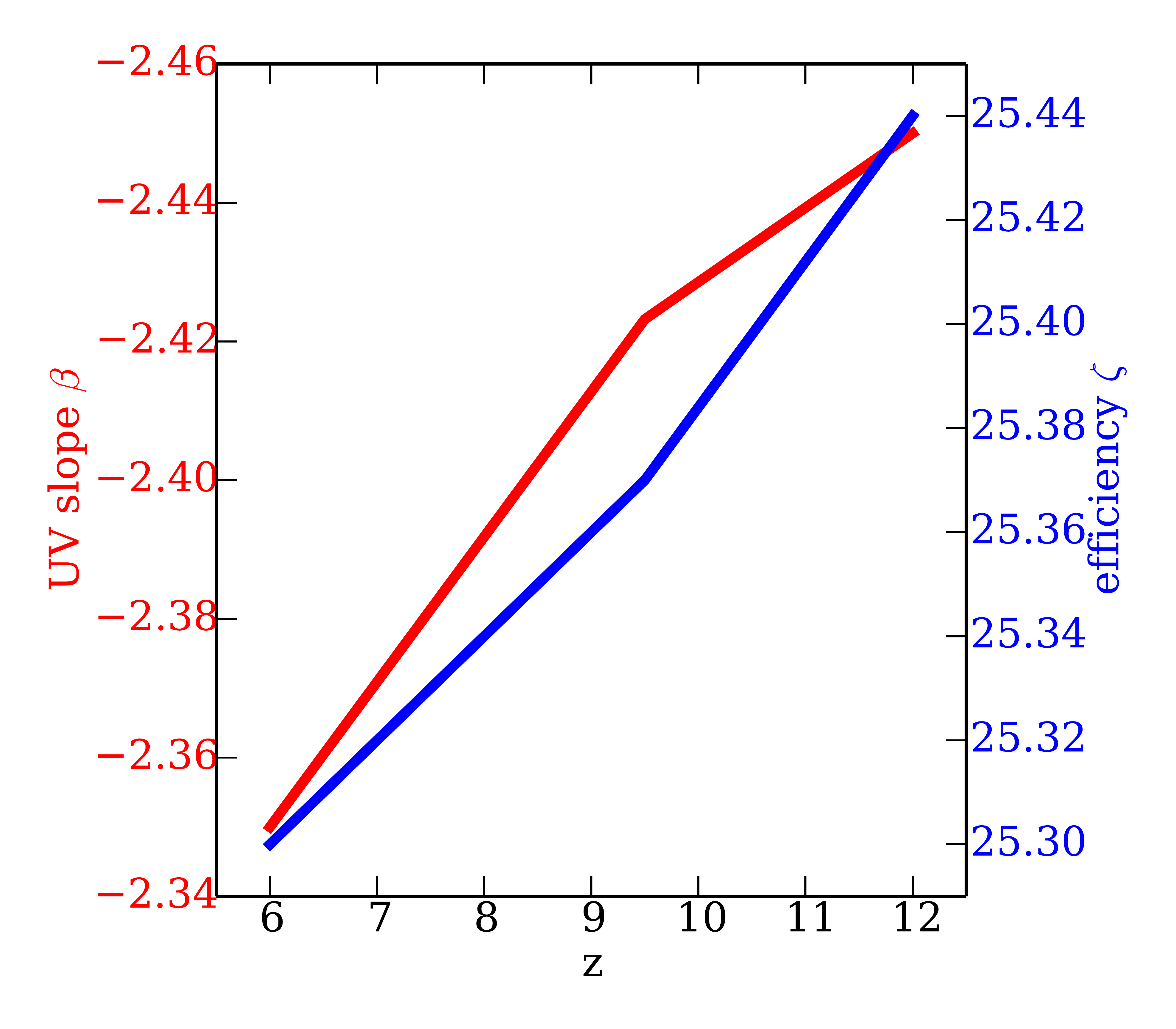}
	 \caption{Evolution of the UV slope (red) and the efficiency (blue).
	 The UV spectrum becomes harder at higher redshifts due to lower metallicities and younger populations.}
	  \label{fig:betazetaevo}
\end{figure}	

The scaling relation between the luminosity at 1500 $\am$ and the galaxy stellar mass is well-established at high redshifts \citep{Stark13, Song16}.
It follows the general trend that more massive galaxies are also brighter in the UV. 
However, there are deviations and the scatter depends on other galaxy properties like the SSFR.
Using the large set of simulated galaxies in the FirstLight database, we can study in detail the typical scatter and the evolution of this relation with redshift.

The top panel of \Fig{MsMUVSSFR} shows the relation at $z\simeq6$, extending from $M_{1500}=-22$ to -10. 
Most of the sample ($M_{1500} \leq -16$) will be observed by ultradeep observations with JWST, assuming a limiting magnitude of ${\rm m}_{\rm lim}=31$ and a uniform K-correction of -2 magnitudes.
See \cite{OShea15} for more details.
This gives us an idea of the observability of the FirstLight galaxies. An extended analysis of mock JWST fields will be described in future papers.

FirstLight is able to reproduce the observed mean and scatter \citep{Song16},
although these observations use a narrower filter with $\Delta \lambda=100$ \am \ centered at 1500 $\AA$\am.
As shown in \Fig{SEDs}, this has a minor effect in the estimation of M$_{1500}$ even for very steep UV spectra.

In Paper I, we estimate the UV magnitude using its correlation with SFR \citep{Madau98}. 
The relation found using the first 15 FirstLight tests agrees well with the average relation using the full sample.
Now, we can study the scatter around this average and its dependencies in more detail. 
At a fixed UV magnitude, the stellar mass may vary by up to an order of magnitude. 
This scatter does not depend significantly on the virial mass (Appendix). 
It is mostly depends on the SSFR, due to the bursty SF history discussed in Paper II.
The scatter in the star forming main sequence sets the scatter in the $\Ms-M_{1500}$ relation. 
Galaxies experience frequent starbursts, every 100 Myr on average at $z\simeq6$. These starbursts with higher-than-average SSFR are brighter in the UV than other quiescent galaxies with lower SSFR but with a similar mass, as shown in \Fig{SEDs}.  

We have seen that the UV magnitude depends on the SSFR at a fixed stellar mass. 
At redshifts higher than 6, the SSFR is typically higher due to higher gas accretion rates (Paper II). 
Therefore, galaxies of the same mass are brighter in  the UV at higher redshifts (\Fig{MsMUVSSFR}).
We see that trend up to $z=12$, when galaxies are up to 2 magnitudes brighter than at $z=6$.
This has important implications for the observability of these extreme star-forming galaxies at these high redshifts.
Due to this increase in absolute luminosity, they may be brighter and easier to detect than their counterparts at lower redshifts.
Future studies will explore the detectability of galaxies at $z\geq10$ in more detail. 

\subsection{Evolution of the UV spectrum}
\label{sec:UV}

The evolution of the UV spectrum of the first galaxies is very relevant for modelling reionization. 
Not only the normalization but also the shape of the spectrum near the Lyman continuum (LyC) limit gives clues about the galactic sources of reionization. Therefore
we define the UV slope $\beta$ as
\begin{equation} 
L_\lambda \propto \lambda^\beta,
\end{equation}
where $L_\lambda$ is the monochromatic luminosity in erg/s/$\am$. 
The slope is computed between 1700 and 2200 $\am$ \citep[][and references therein]{Bouwens16b}.
Another important property is the production efficiency of LyC photons per unit UV luminosity at 1500 $\am$,
\begin{equation} 
\zeta = {\rm log} \left (N_{\rm LyC} / L_{1500} \right ),
\end{equation} 
where $N_{\rm LyC}$ is the number of LyC photons per second.

\Fig{betazeta} shows $\beta$ and $\zeta$ versus $M_{1500}$ (or mass) at $z=6$.
For a wide range of luminosities ($ -22 \leq M_{1500} \leq -14$), most of the FirstLight galaxies show a relatively narrow range of values: $-2.5 \leq \beta \leq -2.2$ and $25 \leq \zeta \leq 25.5$. 
In general, these values show a steep UV slope with a high production of LyC photons. This is the combined effect of low metallicities, young stellar populations and binarity \citep{Ma15}.

The UV slope is consistent with \cite{Bouwens14} observations for  $M_{1500}>-20$.
For brighter galaxies, the observed relation clearly deviates from the above values, reaching $\beta\simeq-1.8$ at $M_{1500}\simeq -22$.
This deviation can be easily explained by the effect of dust extinction on the UV spectra. Dust starts to be relevant in massive galaxies with high dust column densities. 
Future analysis using radiative transfer tools will clarify this issue. In the meanwhile, our values should be compared with extinction-corrected observations.

The production efficiency of LyC photons ($\zeta$) is also consistent with extinction-corrected observations for $M_{1500}<-20$  \citep{Bouwens16c}.
The mean efficiency slightly increases at lower luminosities to $\zeta=25.5$ due to lower metallicities in fainter galaxies. However, observations
 show much higher values of $\zeta=25.75$.
These observations of the faintest bin can be biased toward extreme cases with higher-than-average values.
Future observations of a representative sample of faint galaxies will clarify possible tensions with theory.

A sample of FirstLight galaxies outside this main sequence is observed at lower luminosities, $M_{1500}>-16$, and low masses.
They have much lower values of $\beta$ and $\zeta$, indicating a more quiescent stellar, post-starburst population with lower than average SSFR (Paper II). 	

The mean value of $|\beta |$ and $\zeta$ slowly increases with redshift (\Fig{betazetaevo}).
This trend is expected as the stellar population has lower metallicities at higher redshifts. 
In addition, the time-scales of the starbursts are much shorter (Paper II) and therefore the stellar population that dominates the UV  is younger, yielding a harder UV spectrum.

\subsection{The equivalent width of optical lines}
\label{sec:EW}

\begin{figure}
	\includegraphics[width=\columnwidth]{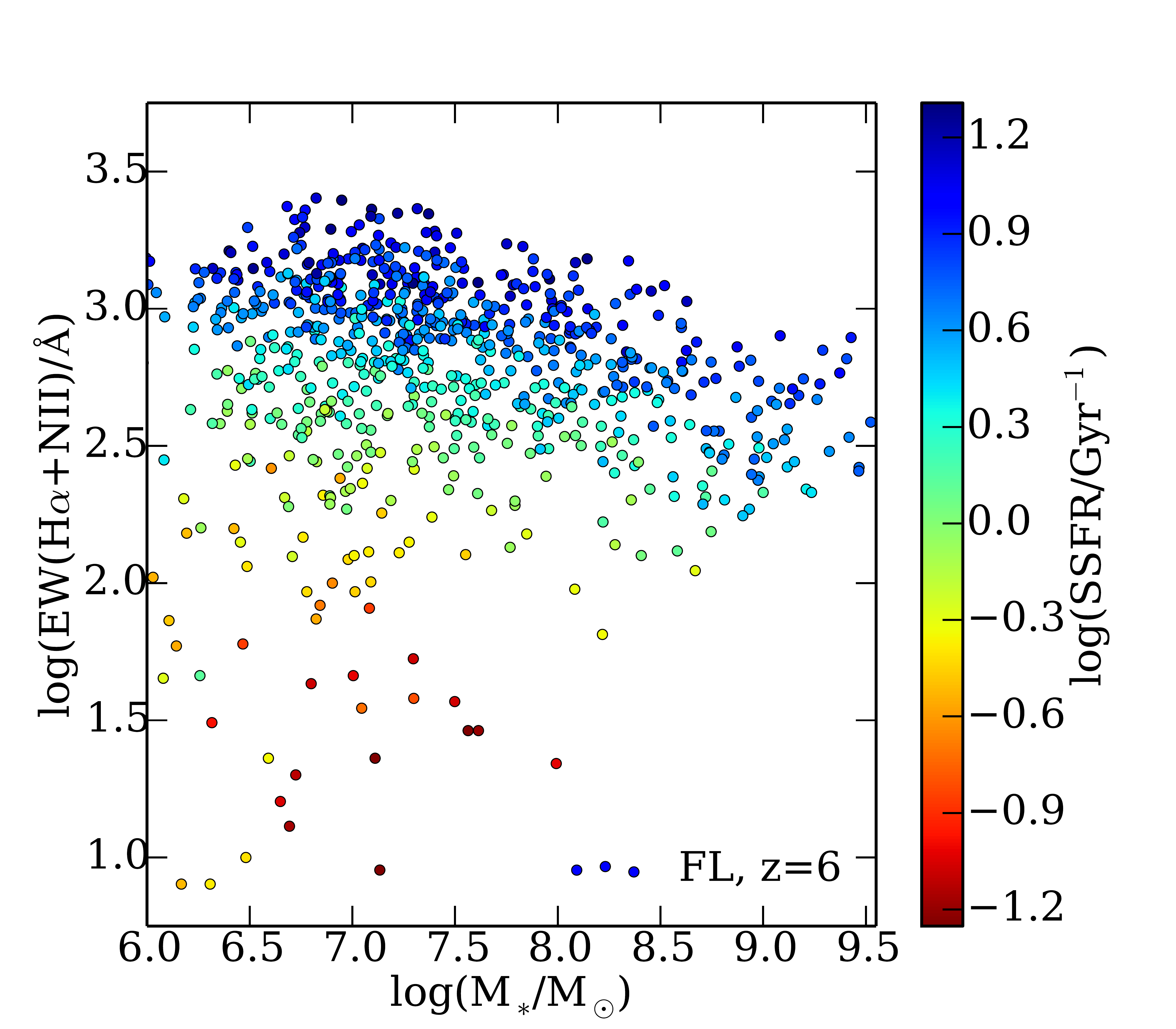}
	 \caption{Equivalent width of the combined H$\alpha$+NII lines versus stellar mass, coloured by SSFR.
	 High EW are common at $z=6$ in star-forming galaxies.} 
	  \label{fig:EWHaNII}
\end{figure}

Among all the optical lines, H$\alpha$ (6563 $\am$)  is the most prominent one. 
It could be combined with the NII doublet (6548 $\am$ and 6584 $\am$) and used as a SFR indicator for galaxies at cosmic dawn.
The equivalent width of the combined lines H$\alpha$+NII are computed from the SEDs, using the line luminosity and the stellar+nebular continuum at both sides of the lines considered.
High EW (300-3000 $\am$) in H$\alpha$+NII are very common at $z=6$ (\Fig{EWHaNII}).
The relative strength of the lines correlates well with the SSFR as expected.

The EW-$\Ms$ relation resembles the SSFR-$\Ms$ relation discussed in detailed in Paper II.
There is a main sequence of star forming galaxies and another population of galaxies with much lower EW$ =10 - 200 \am$, well bellow the star-forming sequence. 
They increase the scatter at low masses, $\Ms\leq 10^7 \ \msun$.
As in previous figures, this is the population of more quiescent galaxies with lower-than-average SSFR.  

The main sequence has a tilt at high masses ($\Ms > 10^8 \ \msun$).
The median equivalent width decreases from EW $\simeq900 \am$ for $\Ms\simeq 10^7 \ \msun$ to   EW $\simeq400 \am$ for $\Ms\simeq 10^9 \ \msun$.
This tilt may be related to the shorter duration of bursts in low-mass galaxies (Paper II). This leads to younger populations on average and higher EW. 
Lower metallicites may also be another factor that yields more ionizing photons (higher $\zeta$), and higher EW.

This picture does not significantly change with redshift (Appendix), although the mean SSFR increases (Paper II).
The EW slightly increases with redshift but this evolution is very mild because both continuum and line strength are higher with increasing SSFR.

\subsection{Stellar mass versus absolute magnitude in the V-band}
\label{sec:VMs}

\begin{figure}
	\includegraphics[width=\columnwidth]{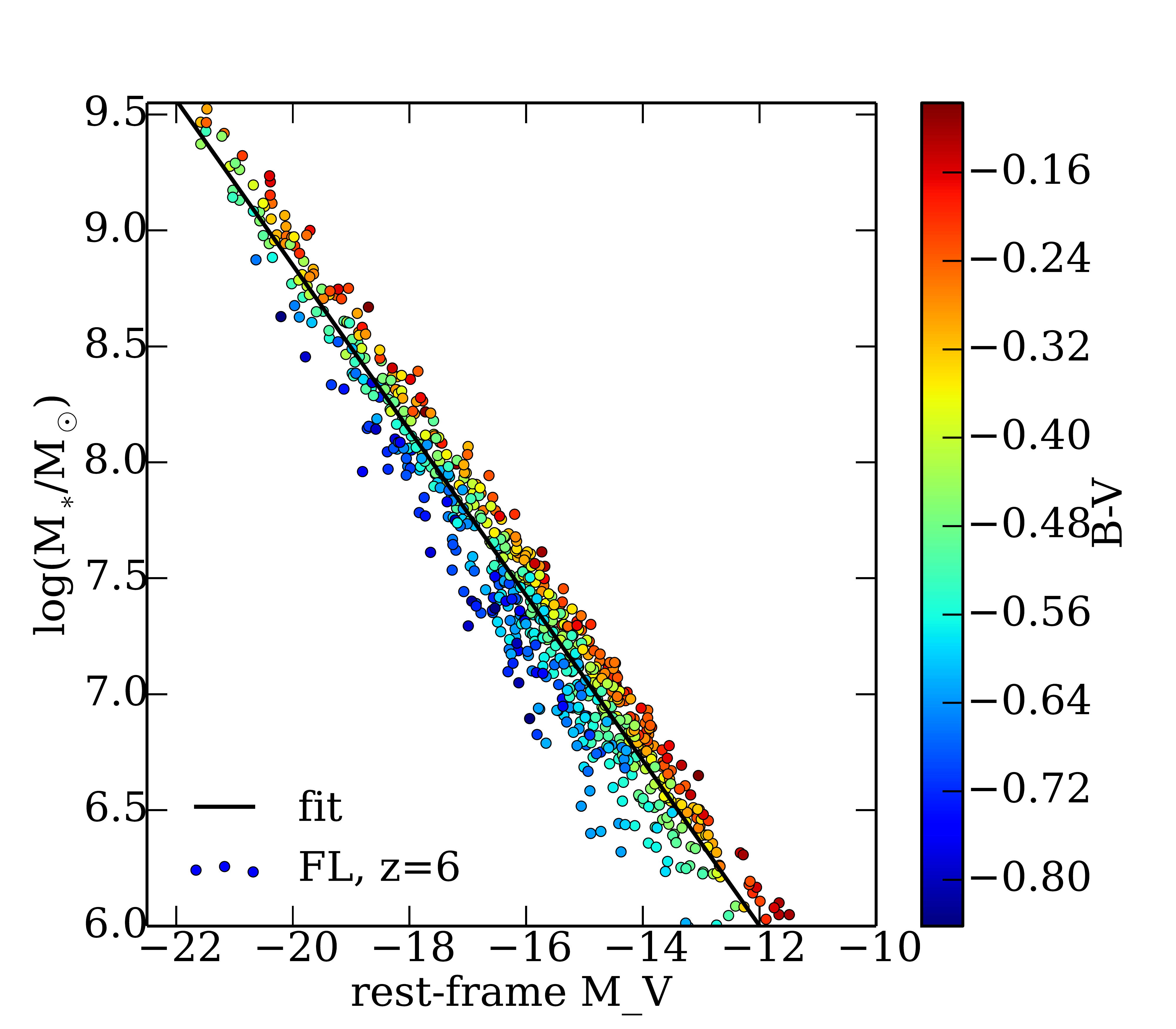}
	 \caption{Stellar mass versus absolute magnitude in the rest-frame V band, coloured by the B-V color.
	 The scatter depends on the B-V color, which correlates with the SSFR.}
	  \label{fig:MVMs}
\end{figure}

In section  \se{MvMs}  we discuss the  $\Ms$-$M_{1500}$ relation. A similar relation exists in the optical part of the SED between 
the stellar mass and the absolute magnitude in the V band (M$\_$V).  \Fig{MVMs}  shows the relation at z=6.
It 
can be described by
 \begin{equation} 
\left( \frac{\Ms}{10^6 \ \msun} \right) = 10^{\alpha_* (M\_{\rm V} - M^*)} ,
\end{equation}
where $ \alpha_* = -0.356 \pm 0.001$, and $M^*=-12.00 \pm 0.01$.
As expected, the scatter in M$\_$V is smaller than in $M_{1500}$ but it is still relevant and increases for fainter galaxies.
The deviation from the main sequence correlates well with the B-V color, such that bluer galaxies have upto 0.5 dex lower masses for the same  V magnitude.
The B-V color strongly depends on the SSFR due to the contamination in the blue band by emission lines (H$\beta$+OIII, see \se{Colors}) 
and therefore the rest-frame optical photometry
not only depends on the galaxy mass, but also on the SFR.
The evolution of this relation with redshift is very similar to the evolution of the $\Ms-M_{1500}$ relation (See Appendix).

\subsection{Colors}
\label{sec:Colors}

\begin{figure}
	\includegraphics[width=\columnwidth]{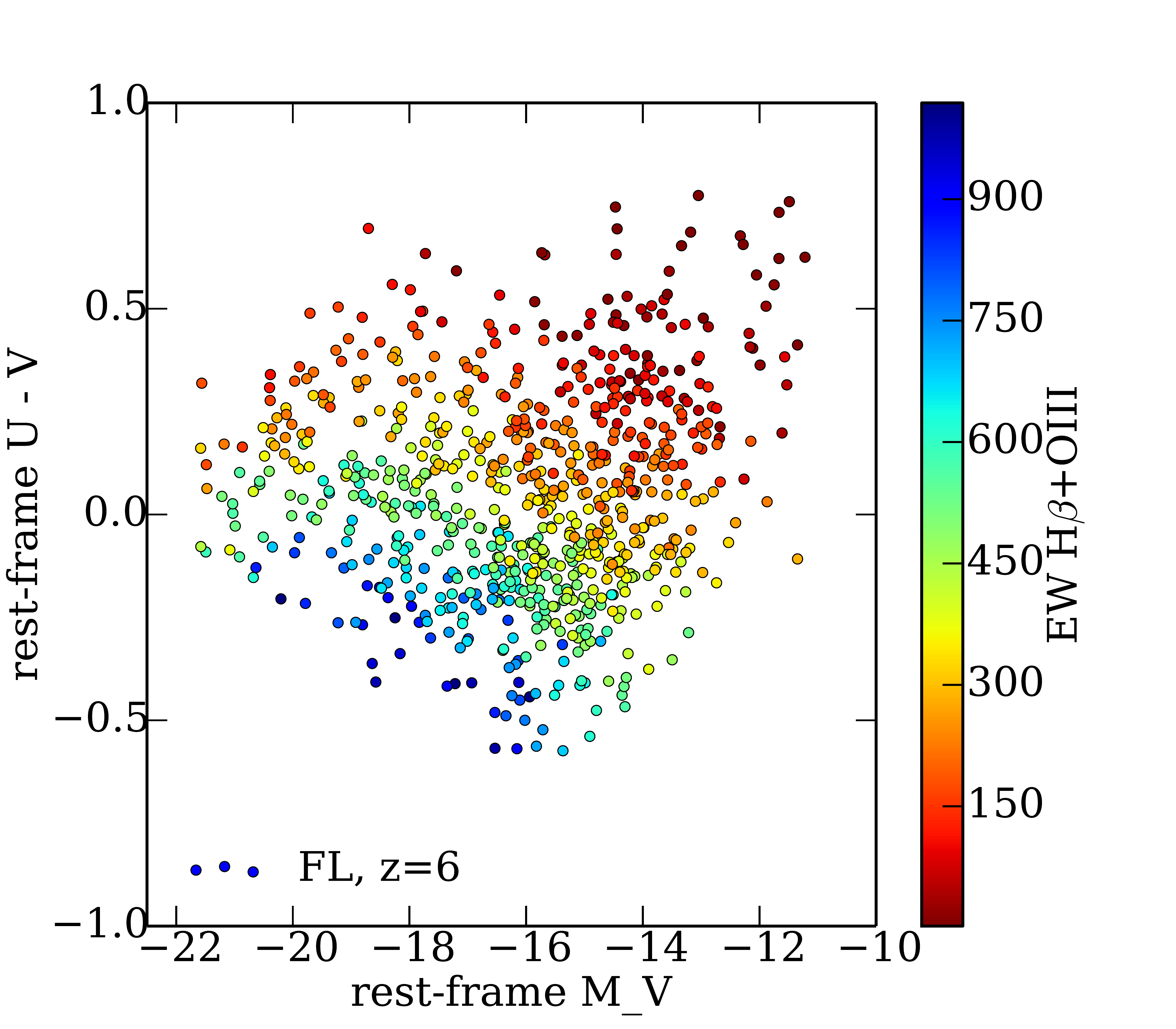}		
	\includegraphics[width=\columnwidth]{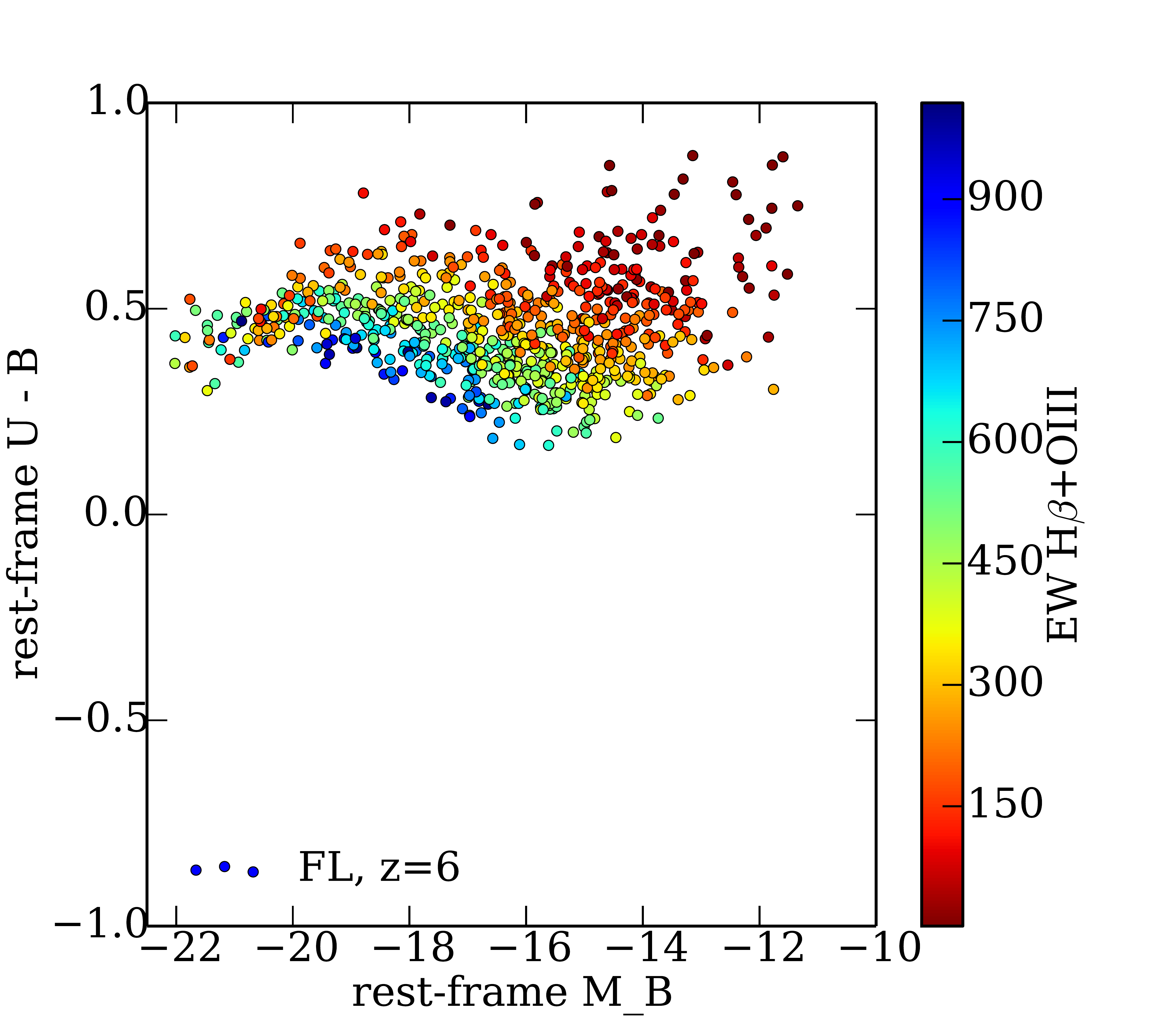}			
		\caption{Color-magnitude plots coloured by the equivalent width of the combined H$\beta$+OIII lines.
		U-V colors show a large diversity of SEDs, from starbursts to almost quiescent galaxies. 
		U-B is contaminated by emission lines.}
	  \label{fig:ColorM1}
\end{figure}

Near-UV to optical colors (U-V)
are commonly used to characterize low-z galaxies. JWST will allow us to make similar color-magnitude diagrams for primeval galaxies.
The rest-frame U-V shows a wide diversity of SEDs at $z=6$ (\Fig{ColorM1}).
Colors vary by more than 1 magnitude and correlate strongly with the EW of the combined H$\beta$+OIII lines or with the SSFR.
The reddest color, U-V $\simeq$0.5, corresponds to the faintest galaxies (M$\_{\rm V} \simeq -12$) with the lowest EW.  
This is the population of low-mass, quiescent galaxies discussed above and in Paper II.
Brighter galaxies, M$\_{\rm V} \simeq -16$, with stellar masses of a few times $10^7 \ \msun$, show the strongest scatter in color.
This is the consequence of the wide diversity of SSFR histories, from almost quiescent galaxies with EW $\simeq$ 100 $\am$ to starbursts with EW $\simeq$ 1000 $\am$.
The brightest galaxies, M$\_{\rm V} \leq -20$, show a smaller scatter in colors (or EW), due to a more continuous and less bursty SSFR histories.
However, these are dust-free colors and dust attenuation may be important for these bright galaxies.

\begin{figure}
	\includegraphics[width=\columnwidth]{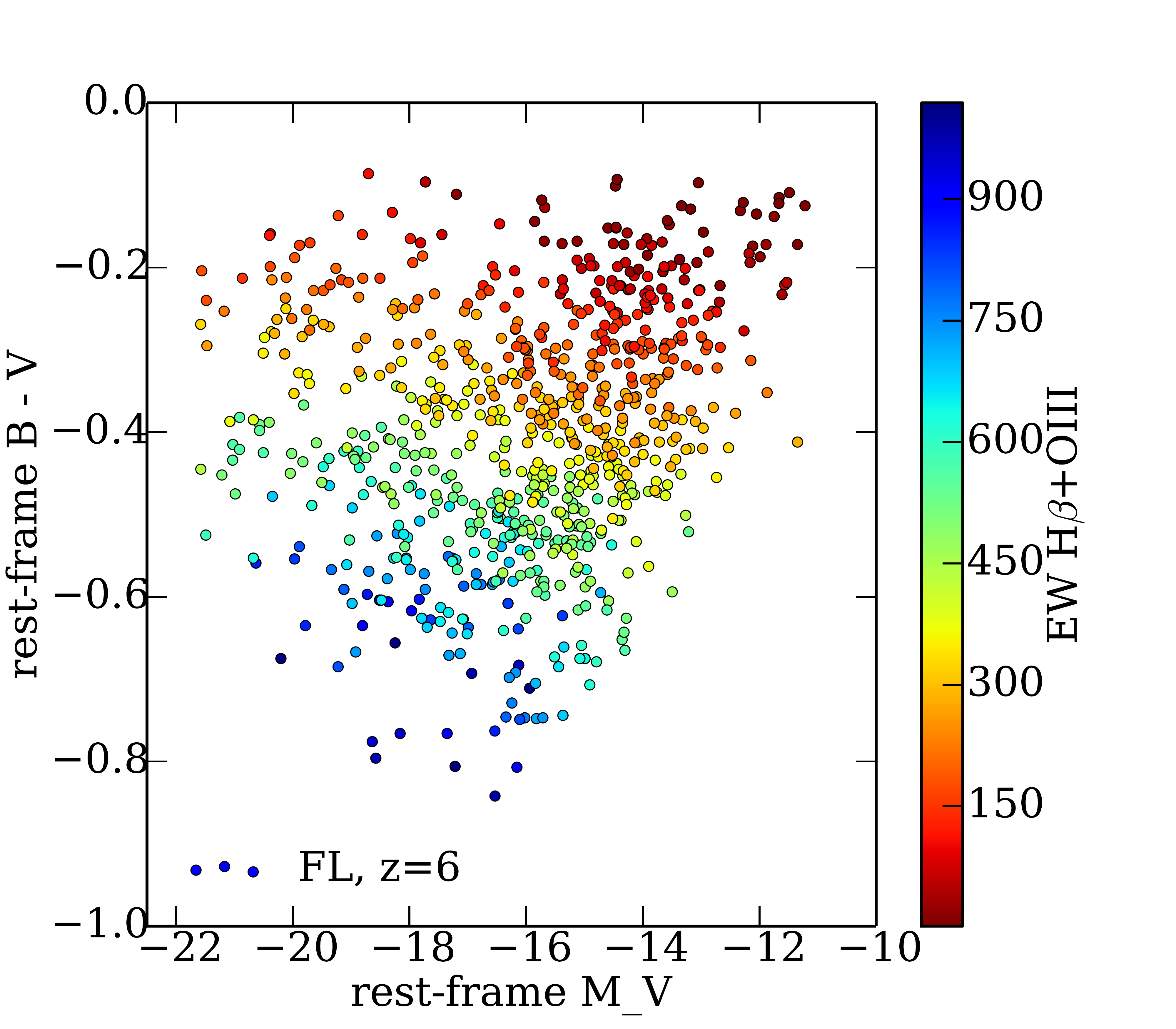}		
	 \caption{B-V color-magnitude plot coloured by the equivalent width of the combined H$\beta$+OIII lines.
	 Due to the line contamination, this color correlates strongly with EW and SSFR.}
	  \label{fig:ColorM2}
\end{figure}

The U-B color (bottom panel of \Fig{ColorM1}) shows a narrow range of values (up to 0.5 magnitude). 
This is due to the contamination by emission lines, particularly  H$\beta$+OIII, discussed above.
These lines increase the luminosity in the B band and complicate the interpretation of the U-B colors.
The contamination of the B band is more evident in the correlation between B-V and EW (\Fig{ColorM2}).
The R-band is also contaminated by H$\alpha$+NII in this case and its colors show a similar behaviour.
Therefore, the B-V and V-R colors can be used as good estimators of the EW of nebular emission lines if there is no spectrum available.

As a cautionary note, the contribution of the emission line to the total flux in a given filter is sensitive to the width of the filter and the shape and relative position of the line within the band. 
This may introduce a dependency on redshift as the line approaches the edge of the filter. 
In practice, for filters with a relatively flat transmission window, strong and broad emission lines may produce a significant redshift effect only when the lines are shifted close to the edge of the filter.

\begin{figure}
	\includegraphics[width=\columnwidth]{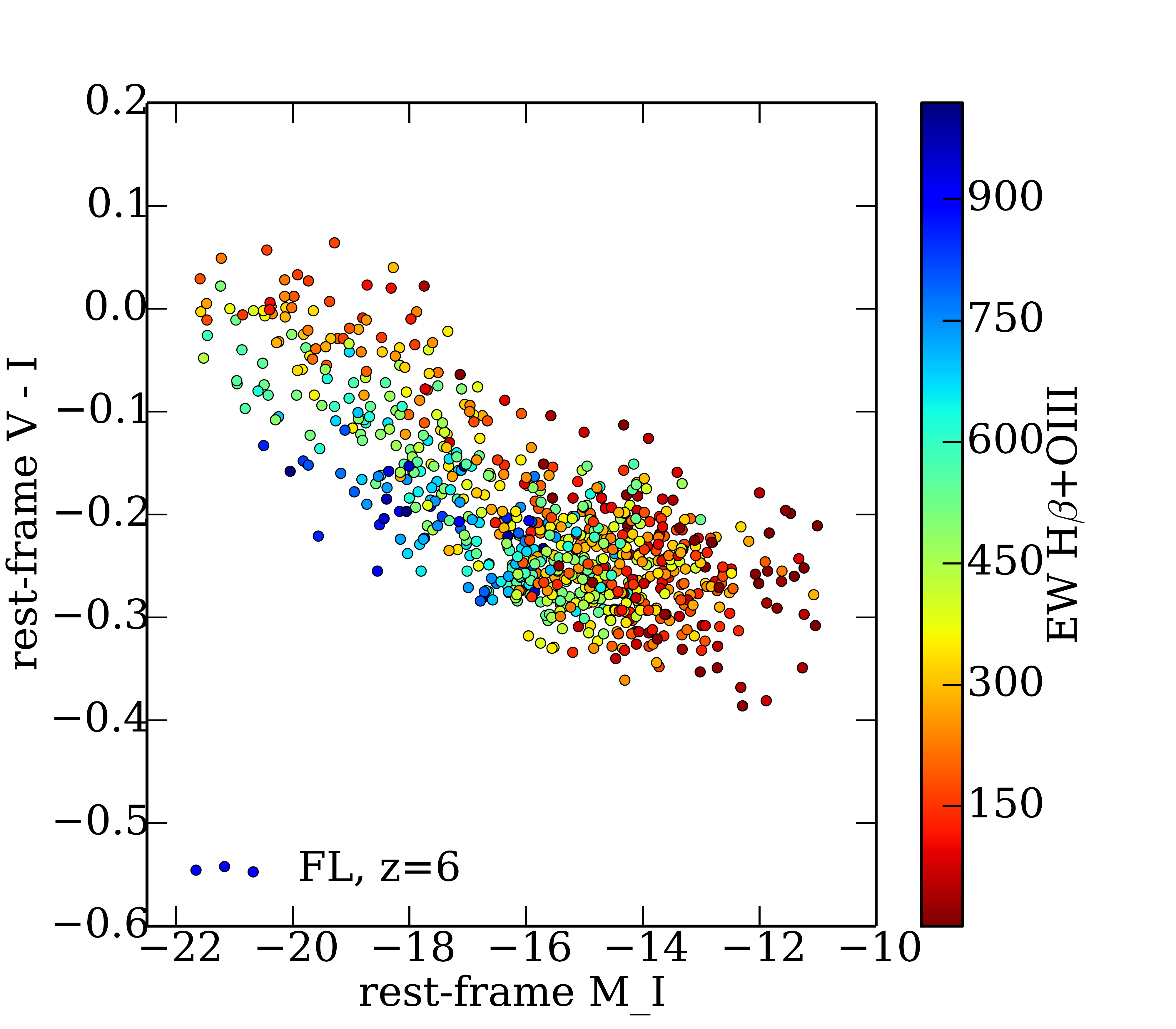}	
	 \caption{V-I color-magnitude plot coloured by the equivalent width of the combined H$\beta$+OIII lines.
	 This trend depends on the evolution of relatively old stellar populations ($\sim$100 Myr).}
	  \label{fig:ColorM3}
\end{figure}

\begin{figure}
	\includegraphics[width=\columnwidth]{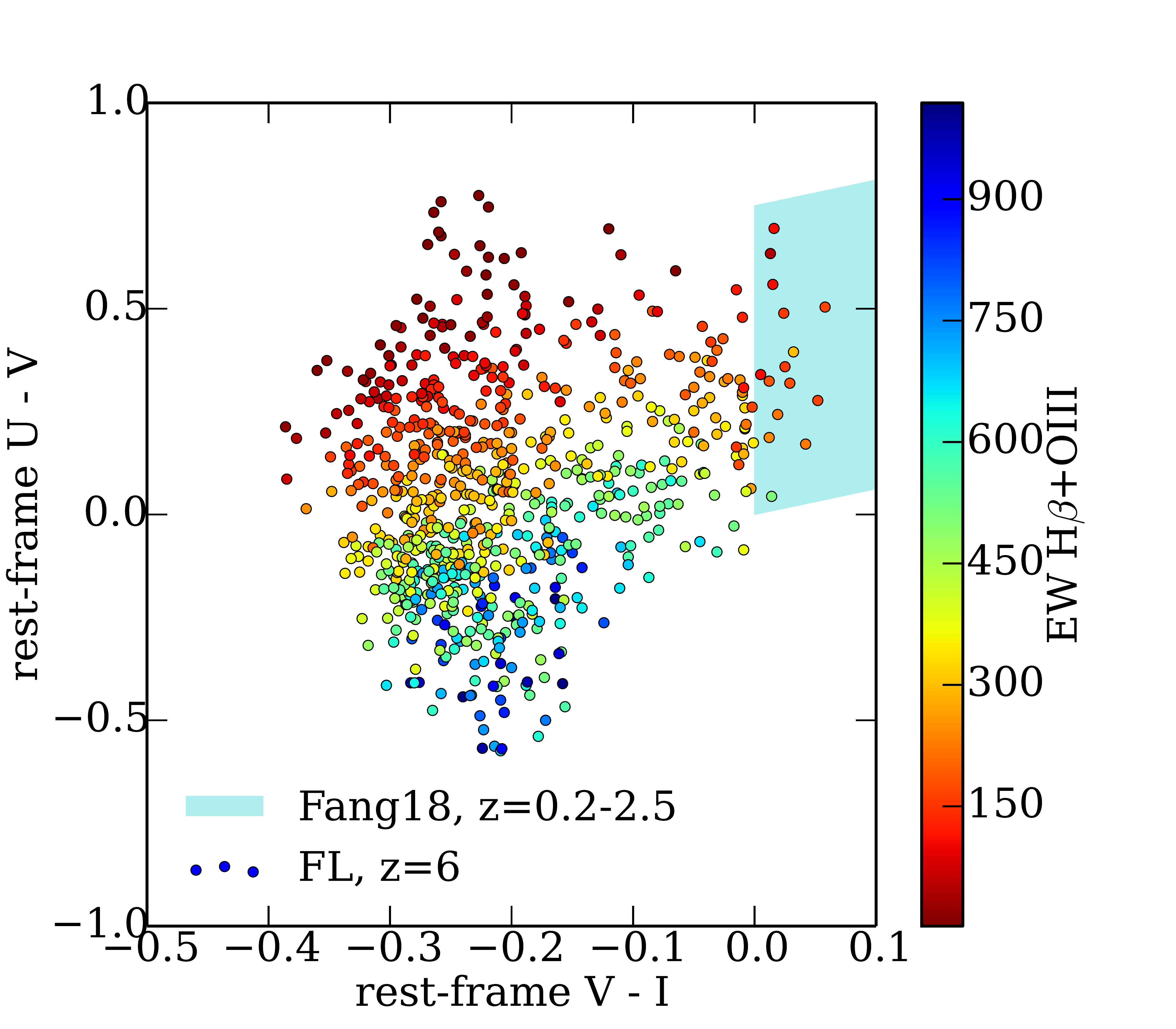}
	 \caption{UVI color-color plot. FirstLight galaxies populate the tip of the blue cloud of star-forming galaxies \citep{Fang18}.} 
	  \label{fig:ColorColor}
\end{figure}	

The V-I color (\Fig{ColorM3}) gives more information about the slope of the galaxy continuum.
FirstLight predicts a flat continuum for relatively bright galaxies, M$\_{\rm I} \leq -18$.
For fainter galaxies, we see a trend of bluer V-I colors. 
This does not correlate with  EW. It depends on the evolution of relatively older stellar populations ($\sim$100 Myr). 
The examples in \Fig{SEDs} show this trend.

Finally, \Fig{ColorColor} shows the U-V versus V-I diagram.
This allow us to compare with commonly used UVJ diagrams, assuming a small I-J color.
FirstLight galaxies at $z=6$ populate the bluest part of the blue cloud of star-forming galaxies. 
The most massive galaxies ($\Ms=1-2 \times 10^9 \ \msun$)
are consistent with observed galaxies at lower redshifts \citep{Fang18}, having V-I $\simeq0$ and U-V $\simeq0.5$. 
Lower mass galaxies follow the observed trend towards bluer V-I colors with an increasing U-V scatter.
As mentioned above, this scatter correlates well with the EW and SSFR.
JWST will be required to observe these regions of the blue cloud.

\subsection{BPT Diagrams}
\label{sec:BPT}
	
\begin{figure}
	\includegraphics[width=\columnwidth]{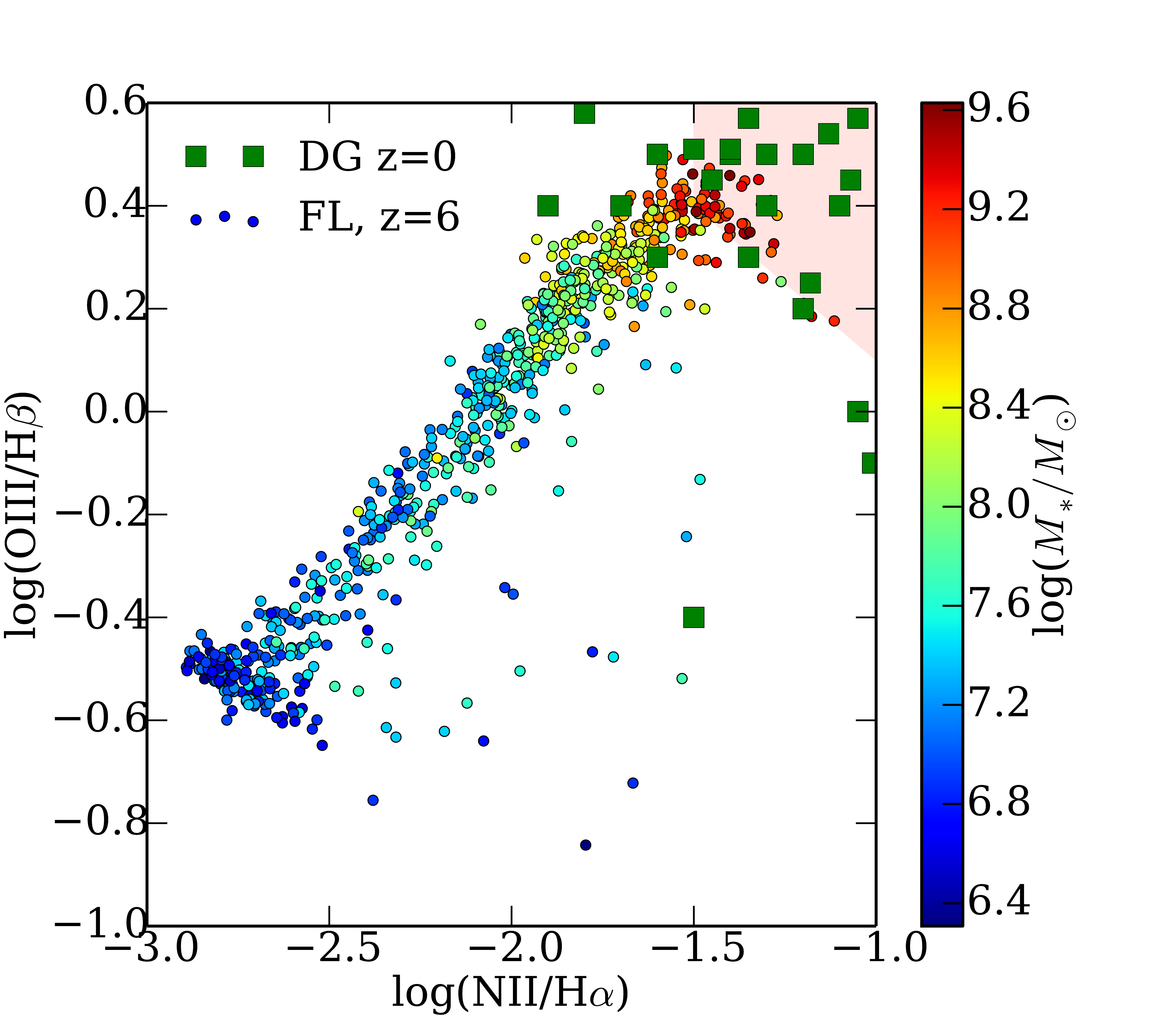}
	\includegraphics[width=\columnwidth]{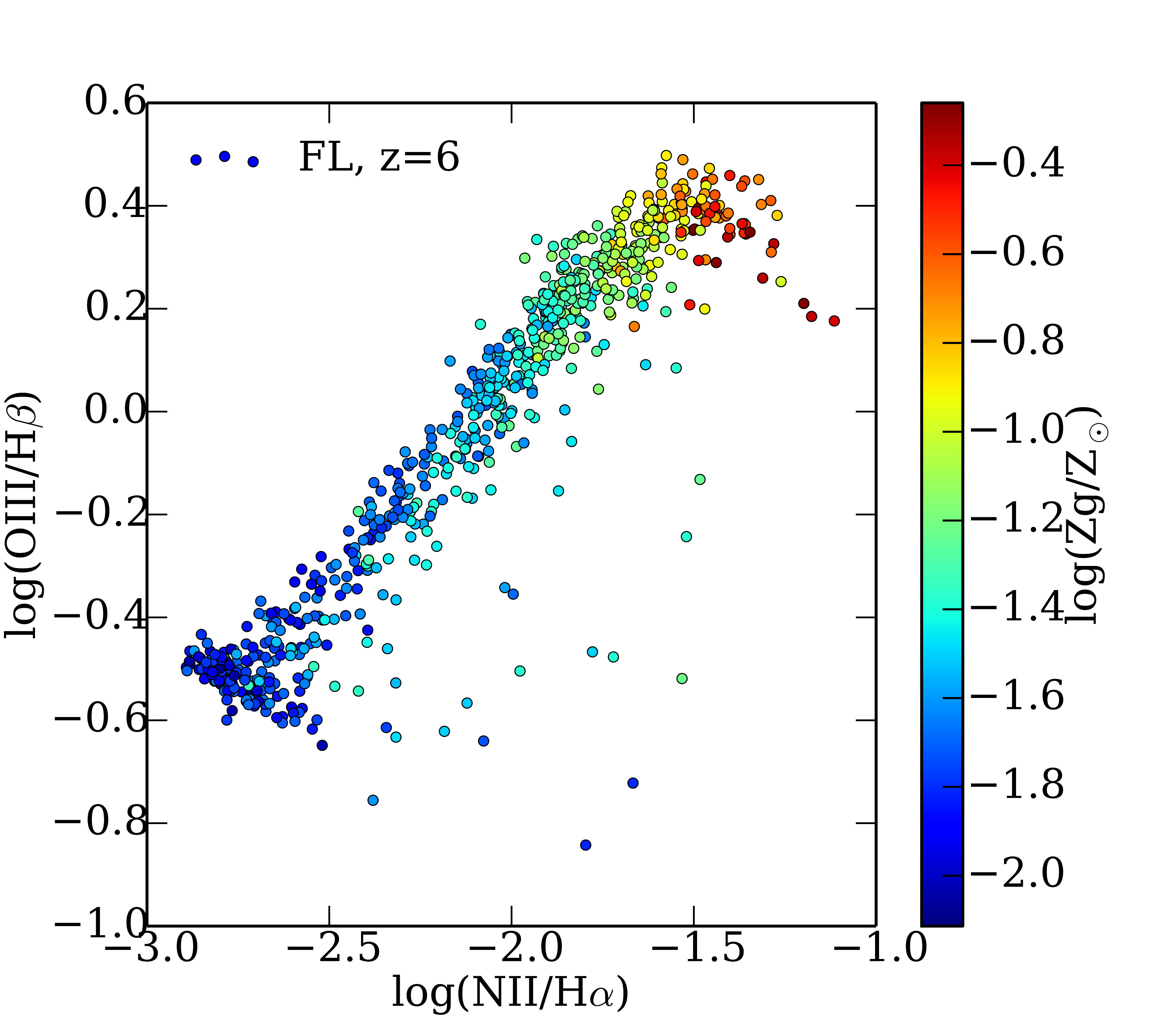}		
	\includegraphics[width=\columnwidth]{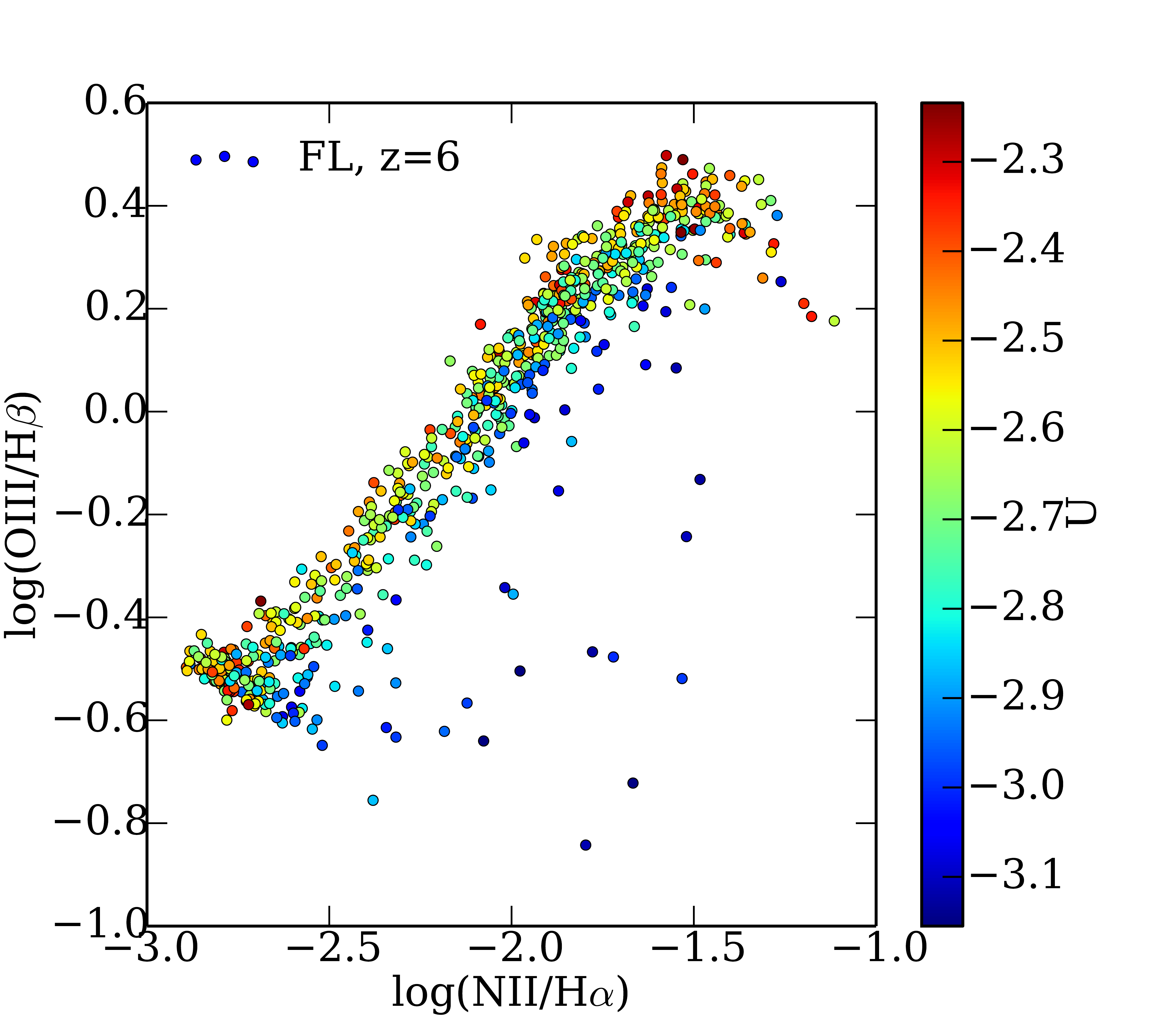}	
	 \caption{BPT diagrams at $z=6$ coloured by stellar mass (top), mean gas metallicity (middle) and mean ionization parameter (bottom). Massive galaxies are consistent with local dwarfs galaxies of similar mass
	 (red region, Kauffmann et al. 2003; green squares, van Zee \& Haynes 2006)
	 Nebular metallicities drive low metal ratios.}
	  \label{fig:BPT}
\end{figure}

The nebular emission lines give some clues about the properties of the HII regions in galaxies using line luminosity ratios like OIII(5007 $\am$)/H$\beta$ and NII(6584 $\am$)/H$\alpha$. In \Fig{BPT} we make predictions of BPT diagrams \citep{BPT}.
The FirstLight galaxies form a clear sequence at $z=6$. 
The most massive galaxies, $\Ms\simeq 1-2 \times 10^9 \ \msun$ occupy the tip of the sequence.
They are consistent with local, low-metallicity dwarfs of similar mass \citep{Kauffmann03}.
High OIII values (OIII/H$\beta\simeq0.5$ dex for NII/H$\alpha \simeq -1.5$ dex)
are similar to the
 line ratios of typical HII regions in low-metallicity dwarf galaxies at $z=0$ \citep{vanZeeHaynes06}. 
These regions have relatively high luminosities in OIII with respect to H$\beta$. 
Therefore, strong OIII-emitters at low-$z$ could be good analogs of galaxies responsible for reionization \citep{ Fletcher18}, 
although none of our galaxies reaches the highest values of  OIII/H$\beta\simeq$ 1 dex found in some of these analogs. 
This is mostly due to the low-metallicity of the primeval galaxies. 

Low-mass galaxies have low luminosity ratios, reaching extremely low values of OIII/H$\beta\simeq -0.5$ dex and NII/H$\alpha \simeq -2.75$ dex for $\Ms\simeq10^6 \ \msun$. 
This is mostly driven by the low nebular metallicities of these low-mass galaxies (middle panel of \Fig{BPT}), which reach log(Z$_{\rm g}$/Z$_\odot)=-2$.  	

The predicted relation is probably too tight due to our assumption of a constant nebular density. 
The ionization parameter of the galaxies along that sequence has a narrow set of values: -2.2 < U < -2.7 (bottom panel of \Fig{BPT}).
For lower values, we see a significant deviation from the main sequence, giving lower OIII/H$\beta$ ratios and higher NII/H$\alpha$ values.
This happens for a small fraction of quiescent galaxies.

\section{Conclusions and Discussion}
\label{sec:summary}

We have used the FirstLight database of zoom-in cosmological simulations (paper I) to study the spectral energy distributions (SEDs) of $\sim$300 distinct galaxies with a stellar mass between $\Ms = 10^6$ and $3 \times 10^9 \msun$ during cosmic dawn ($z = 6 - 12$). The main results can be summarized as follows:
\begin{itemize}
\item
There is a large diversity of SEDs, from starbursts to quiescence, even at a fixed stellar mass.
\item
This drives a large scatter in the scaling relation between the luminosity at 1500 $\am$ and stellar mass. 
At a fixed stellar mass, the UV magnitude may vary by up to 1 order of magnitude, driven by variations in the specific star formation rate.
 \item
 This scaling relation evolves with redshift. 
 Galaxies of the same mass are up to 2 magnitudes brighter at $z=12$ than at $z=6$.
 \item
 The absolute value of the UV slope and the production efficiency of Lyman continuum photons are high, consistent with dust-corrected observations.
 \item
 The equivalent width of optical nebular lines, like H$\alpha$+NII are typically high, of the order of 1000 $\am$, in SF galaxies at high-$z$.
 \item
 U-V color-magnitude relations show a 1 magnitude scatter in color due to the diversity of galaxies and it correlates strongly with SSFR and EW of nebular lines. 
 \item
 Color-color diagrams show that galaxies at $z=6$ populate the bluest end of the blue cloud of star-forming galaxies.
 \item
 BPT diagrams show high OIII/H$\beta$ ratios  for the brightest galaxies with stellar masses of $\sim10^9 \msun$. 
 However, the lower metallicity of fainter galaxies drives the population towards low metal-line ratios.   
\end{itemize}

One of the caveats in the present analysis is the omission of radiative transfer effects from the intervening gas.
We found that dust attenuation may be relevant for the most massive galaxies in the FirstLight sample ($\Ms \geq 10^9 \ \msun$). The low metallicities of smaller galaxies mitigate this issue. 
Future radiative transfer calculations in a post-processing step will clarify the importance of this effect.
In addition, the effect of radiation leakage from the galactic HII regions has also not been included in the analysis and in the simulations.
The addition of a radiative transfer module into the simulations will address this effect in  future simulations.

The calculation of the SEDs of this paper relays on the BPASS spectral synthesis code. 
It includes binary stars which generally lead to better correspondence between synthetic and observed spectra
\citep{Ma16, Rosdahl18}. 
However, this is an evolving field and BPASS has its own limitations. 
For example, it does not include rotating stars. This has been shown to have an effect comparable to the inclusion of binarity \citep{Byler17}. 
Future analysis with more sophisticated  spectral synthesis codes will address possible systematics.

The FirstLight database is publicly available at this website\footnote{\url{http://www.ita.uni-heidelberg.de/~ceverino/FirstLight}}.
It includes all galaxy properties described in Paper II along with the SEDs used in this paper.
It provides a useful set of templates with complex star formation histories consistent with cosmological gas accretion into galaxies.
These templates will be the building blocks for future mock ultradeep fields that will be compared with observations coming from JWST and the next generation of telescopes in the incoming decade.

\section*{Acknowledgements}
We thank Luis Colina and Santiago Arribas for fruitful discussions.
This work has been funded by  the ERC Advanced Grant, STARLIGHT: Formation of the First Stars (project number 339177). DC is a DAWN fellow.
RSK and SCOG also acknowledge support from the DFG via SFB 881 `The Milky Way System' (sub-projects B1, B2 and B8) and SPP 1573 `Physics of the Interstellar Medium' (grant number GL 668/2-1) and KL 1358/19-2.
The authors gratefully acknowledge the Gauss Center for Supercomputing for funding this project by providing computing time on the GCS Supercomputer SuperMUC at Leibniz Supercomputing Centre (Project ID: pr92za).
The authors acknowledge support by the state of Baden-W\"{u}rttemberg through bwHPC.
We thank the BPASS team for sharing their database of SSPs and emission lines. This work made use of the v2.1 of the Binary Population and Spectral Synthesis (BPASS) models as last described in \cite{Eldridge17}. 




\bibliographystyle{mnras}
\bibliography{SED5} 


\appendix

\section{Relations at redshifts $z\geq6$}

\begin{figure}
	\includegraphics[width=\columnwidth]{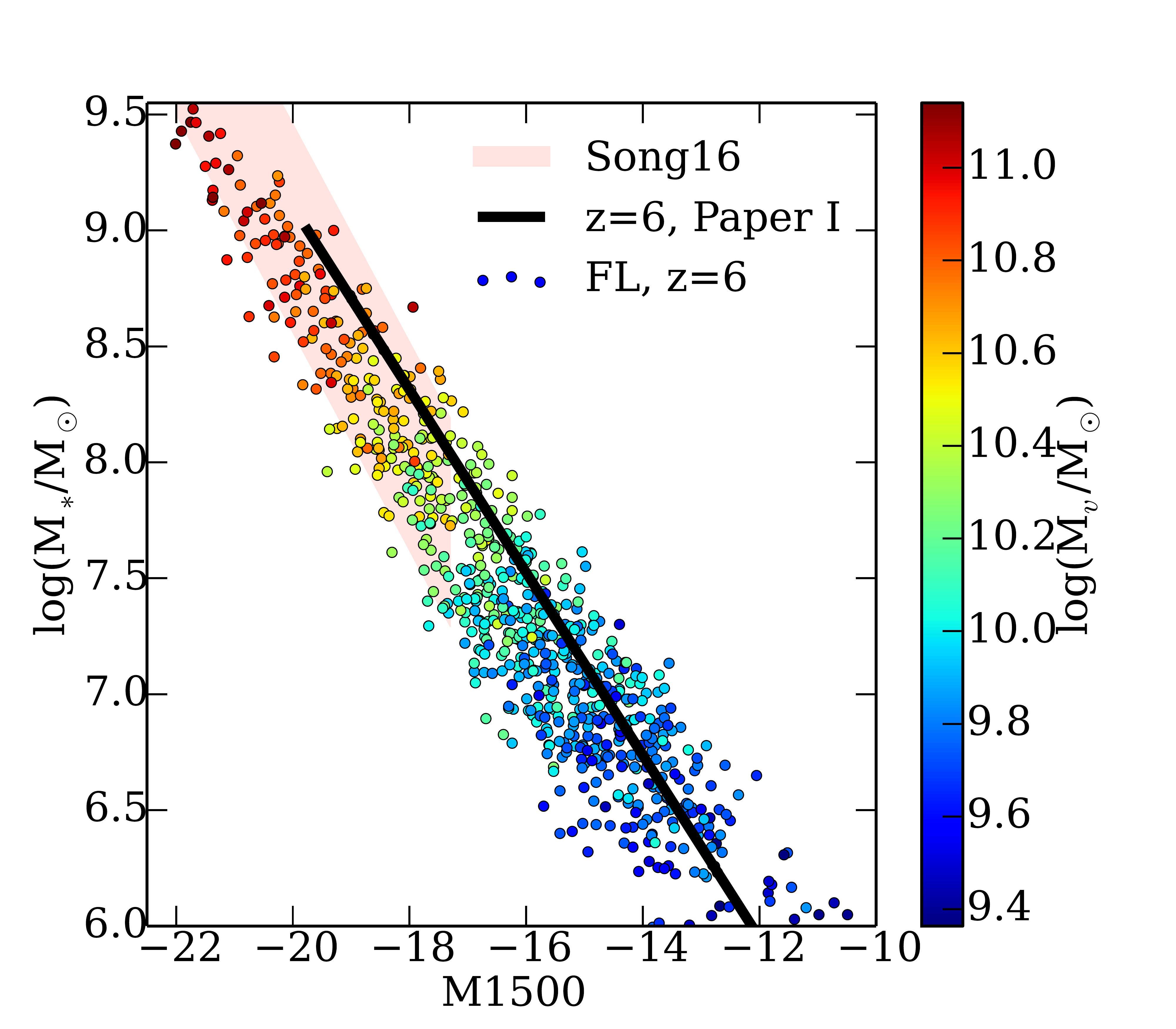}
	\includegraphics[width=\columnwidth]{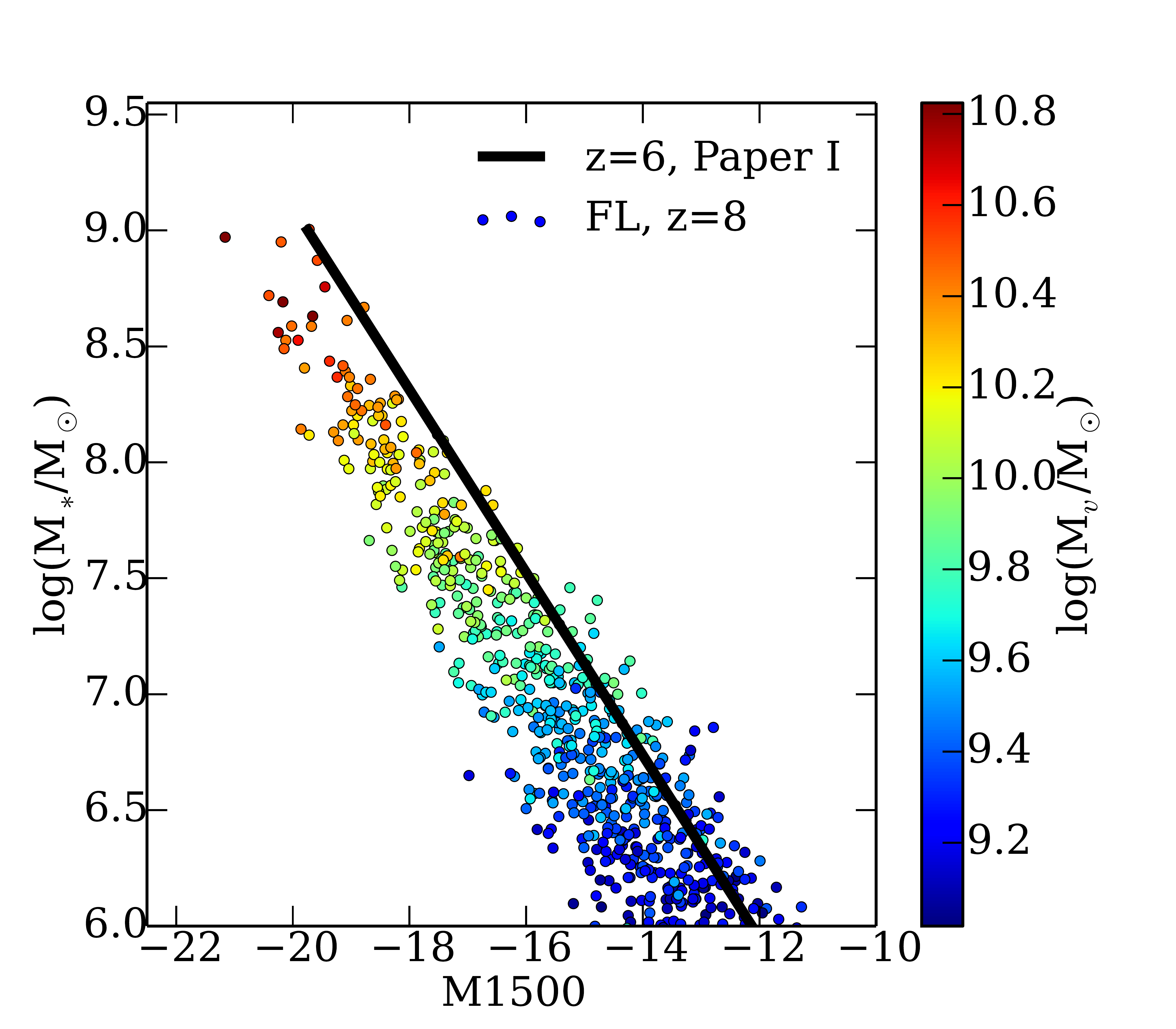}
	\includegraphics[width=\columnwidth]{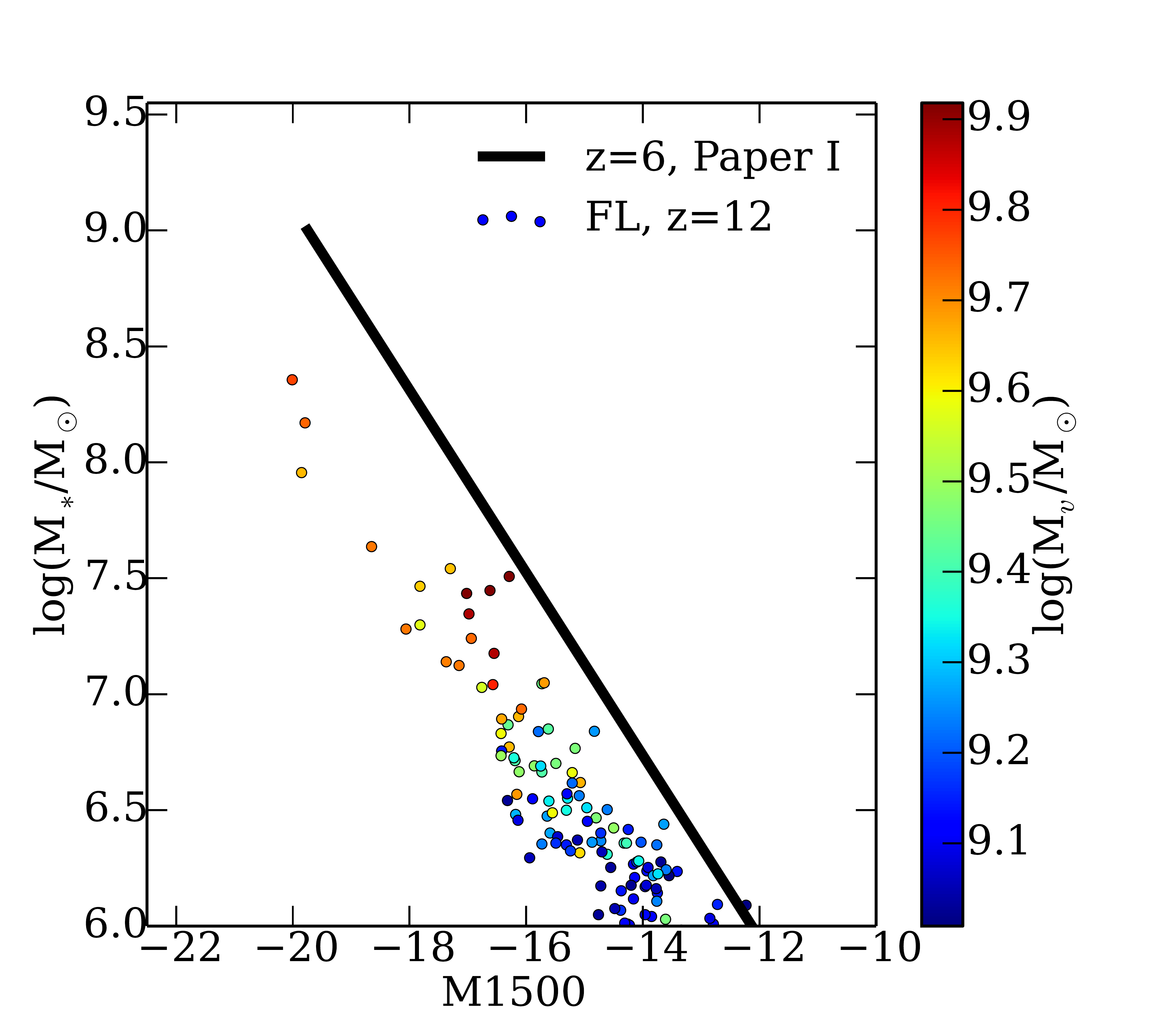}	
	 \caption{Stellar mass versus $M_{1500}$, coloured by virial mass}
	  \label{fig:MsMUVMv}
\end{figure}

In this appendix, we show some of the relations discussed in this paper at higher redshifts, $z=8$ and 12.
\Fig{MsMUVMv} shows the evolution of the scaling relation between the luminosity at 1500 $\am$ and the galaxy stellar mass similar to \Fig{MsMUVSSFR} but now coloured by the virial mass. The scatter at a fixed luminosity is not driven by differences in halo masses. 

\begin{figure}
	\includegraphics[width=\columnwidth]{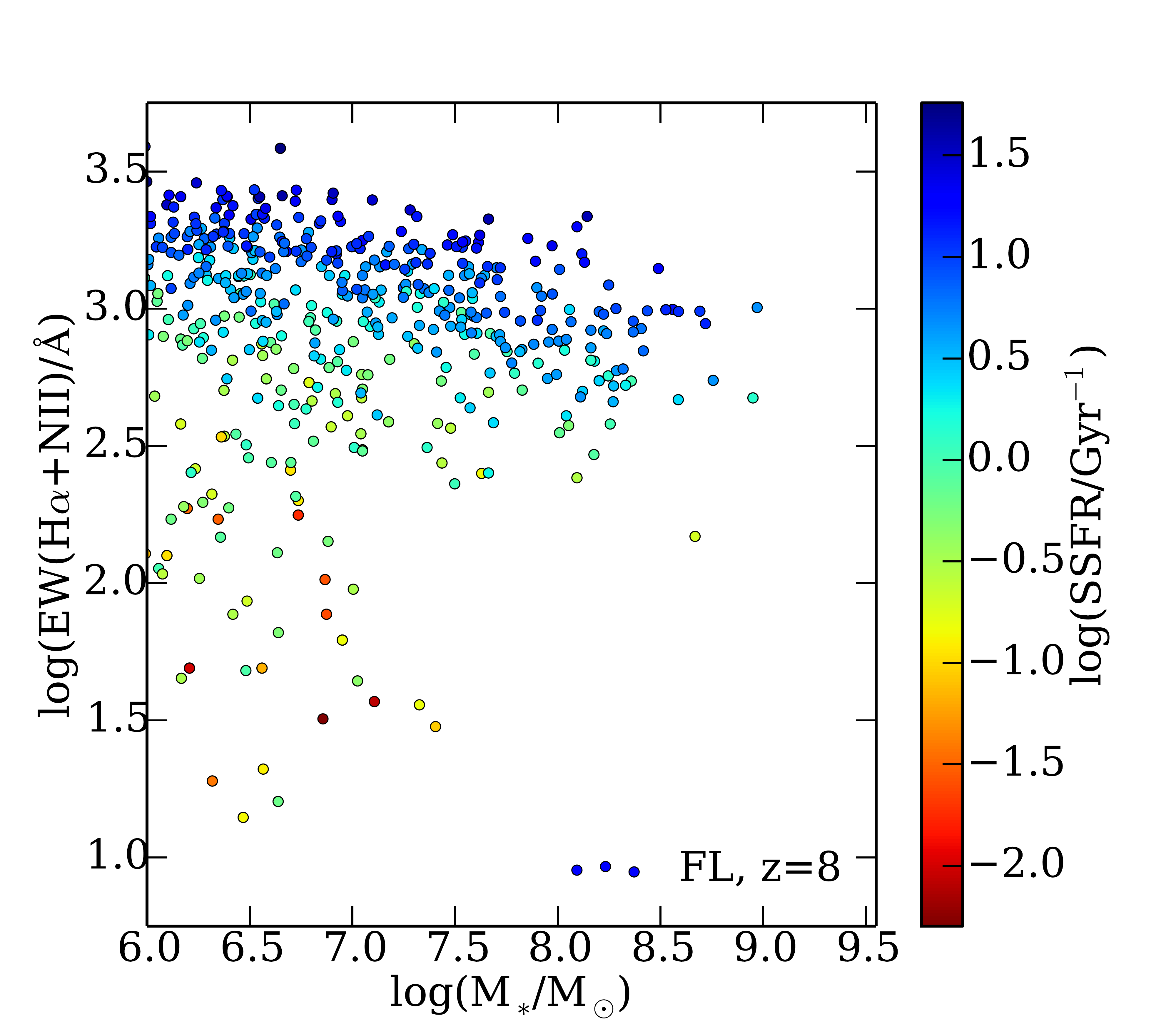}
	\includegraphics[width=\columnwidth]{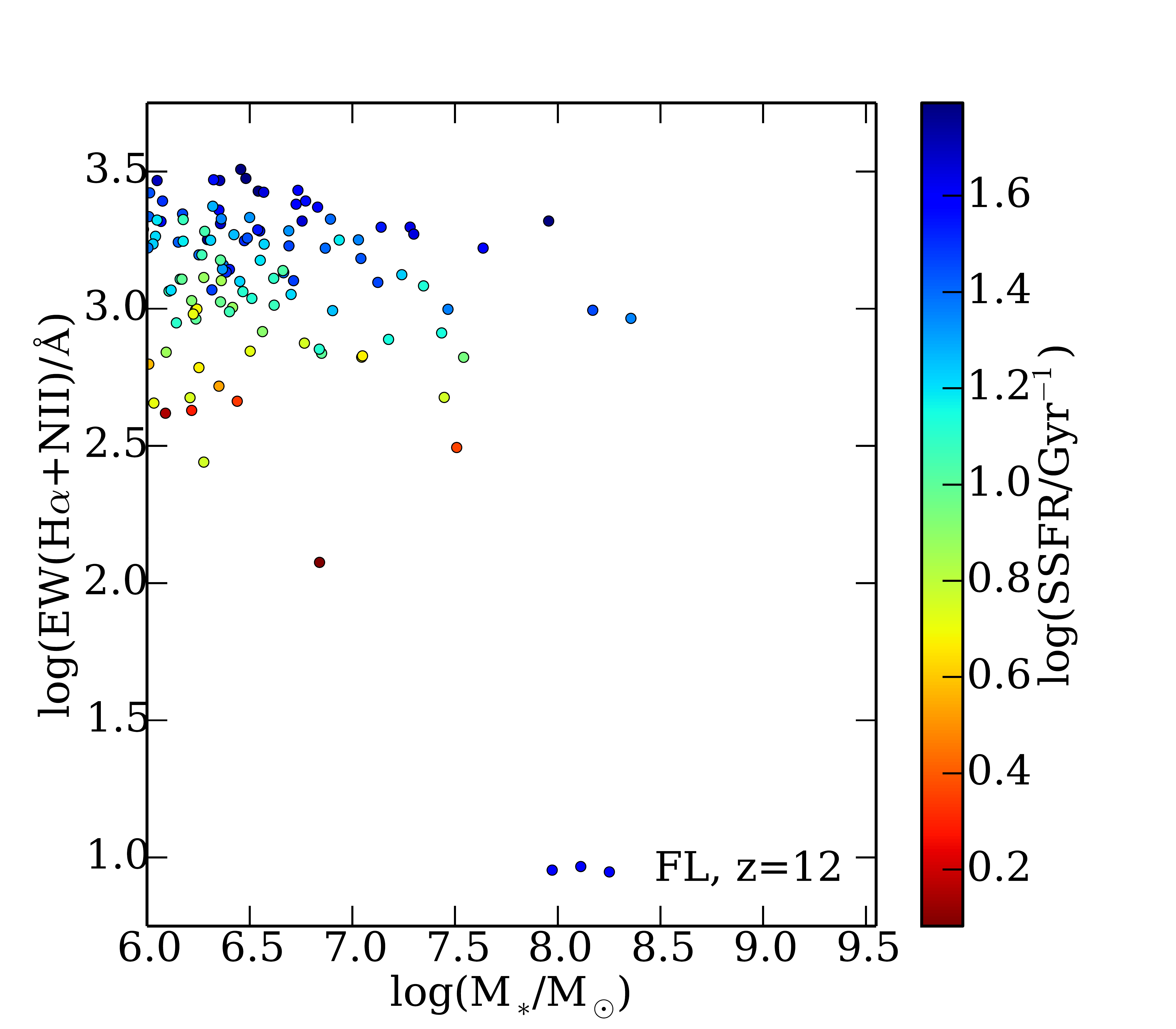}
	 \caption{Equivalent width of the combined H$\alpha$+NII lines versus stellar mass, coloured by SSFR at $z=8$ and 12.} 
	  \label{fig:A2}
\end{figure}

\begin{figure}
	\includegraphics[width=\columnwidth]{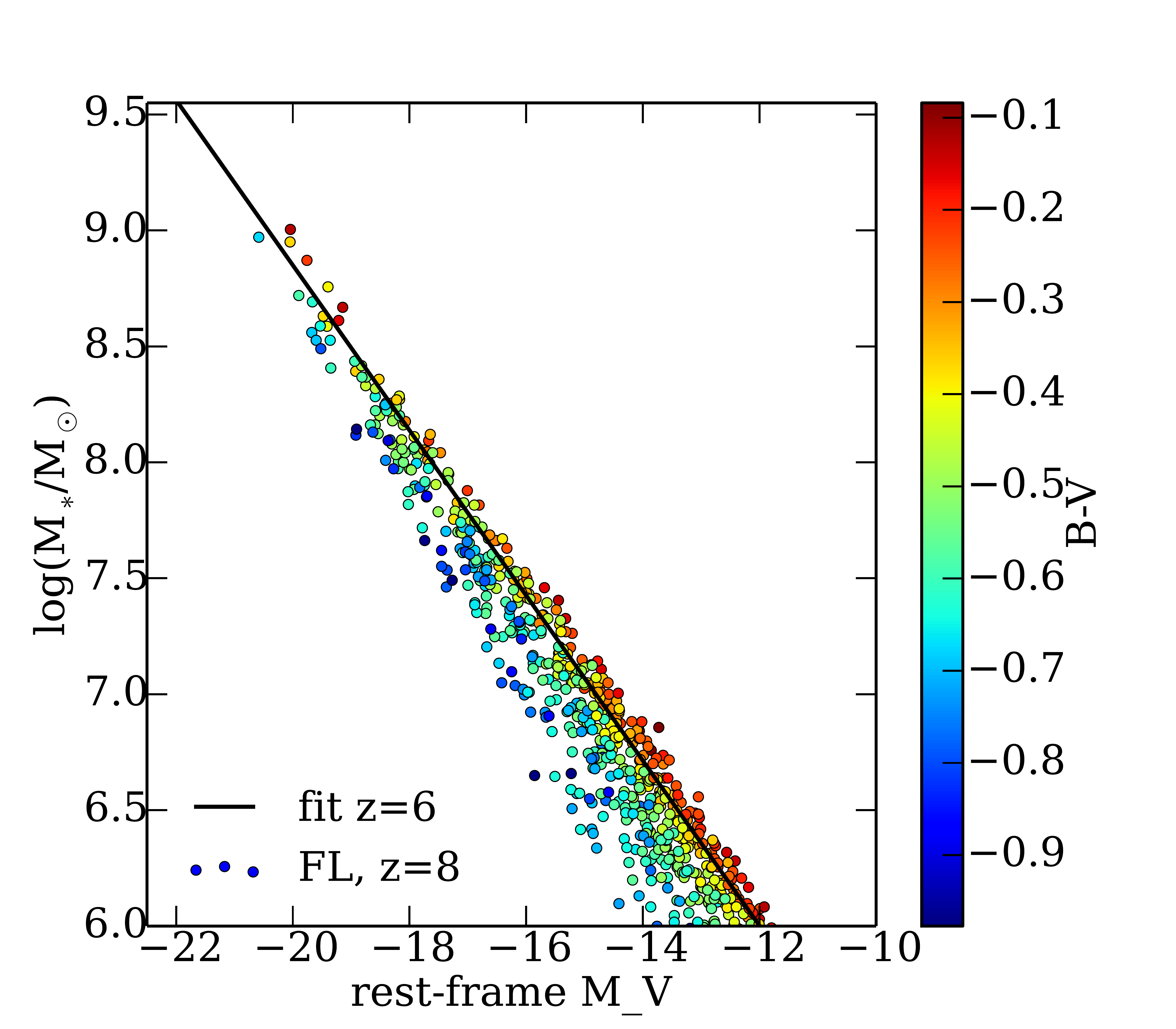}
	\includegraphics[width=\columnwidth]{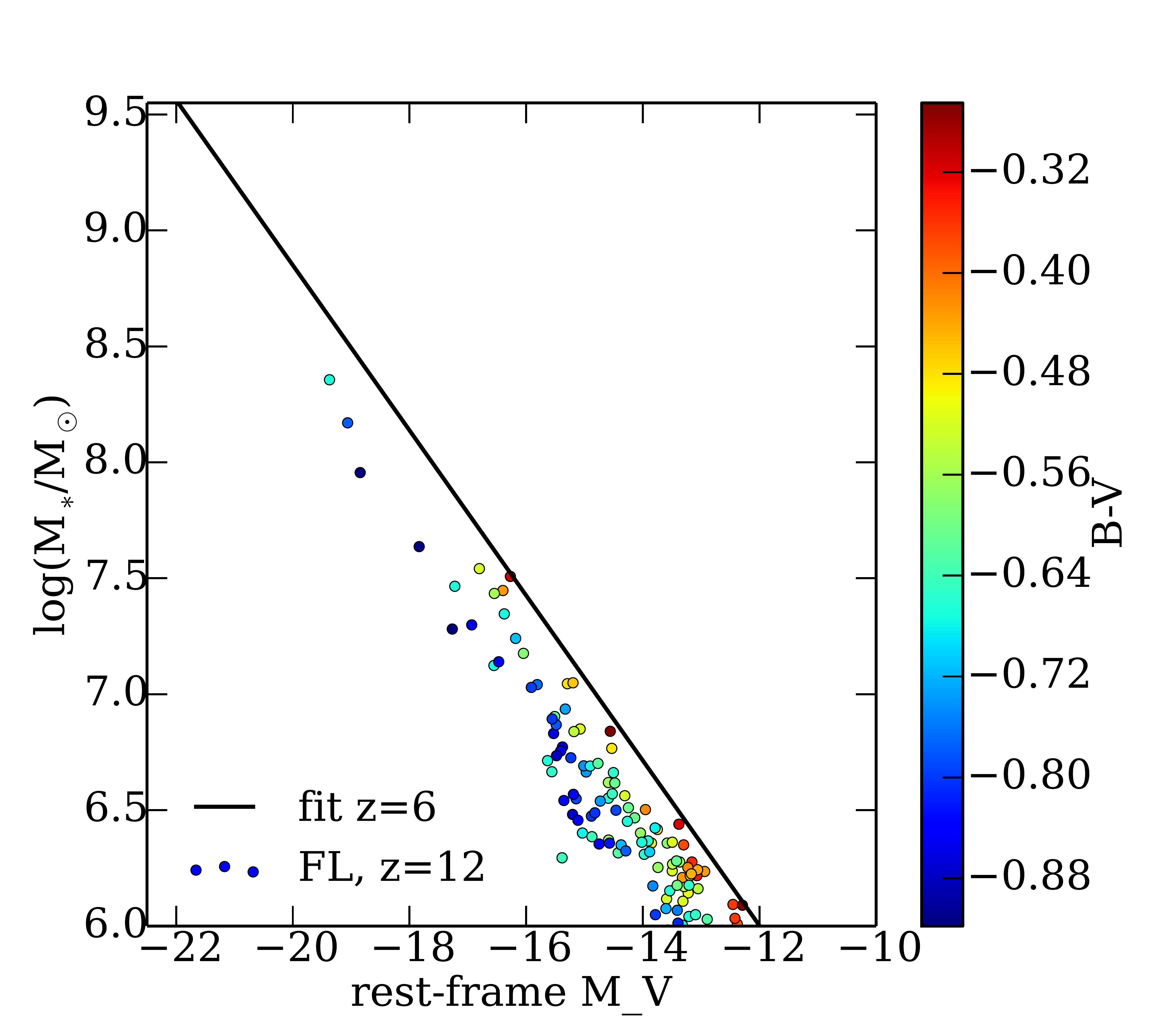}	
	 \caption{Stellar mass versus absolute magnitude in the rest-frame V band, coloured by the B-V color at $z=8$ and 12.} 
	  \label{fig:A3}
\end{figure}

\begin{figure}
	\includegraphics[width=\columnwidth]{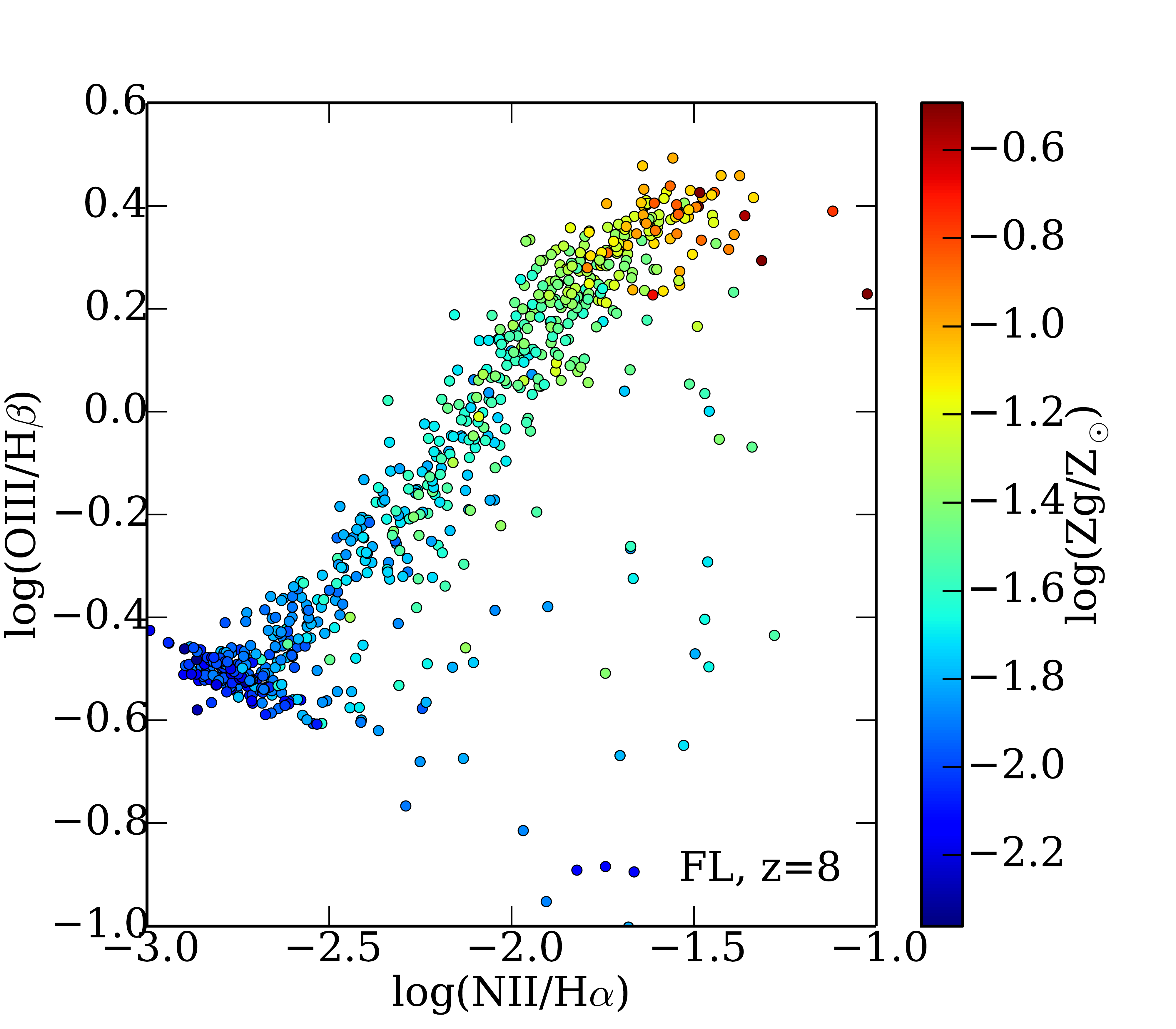}	
	\includegraphics[width=\columnwidth]{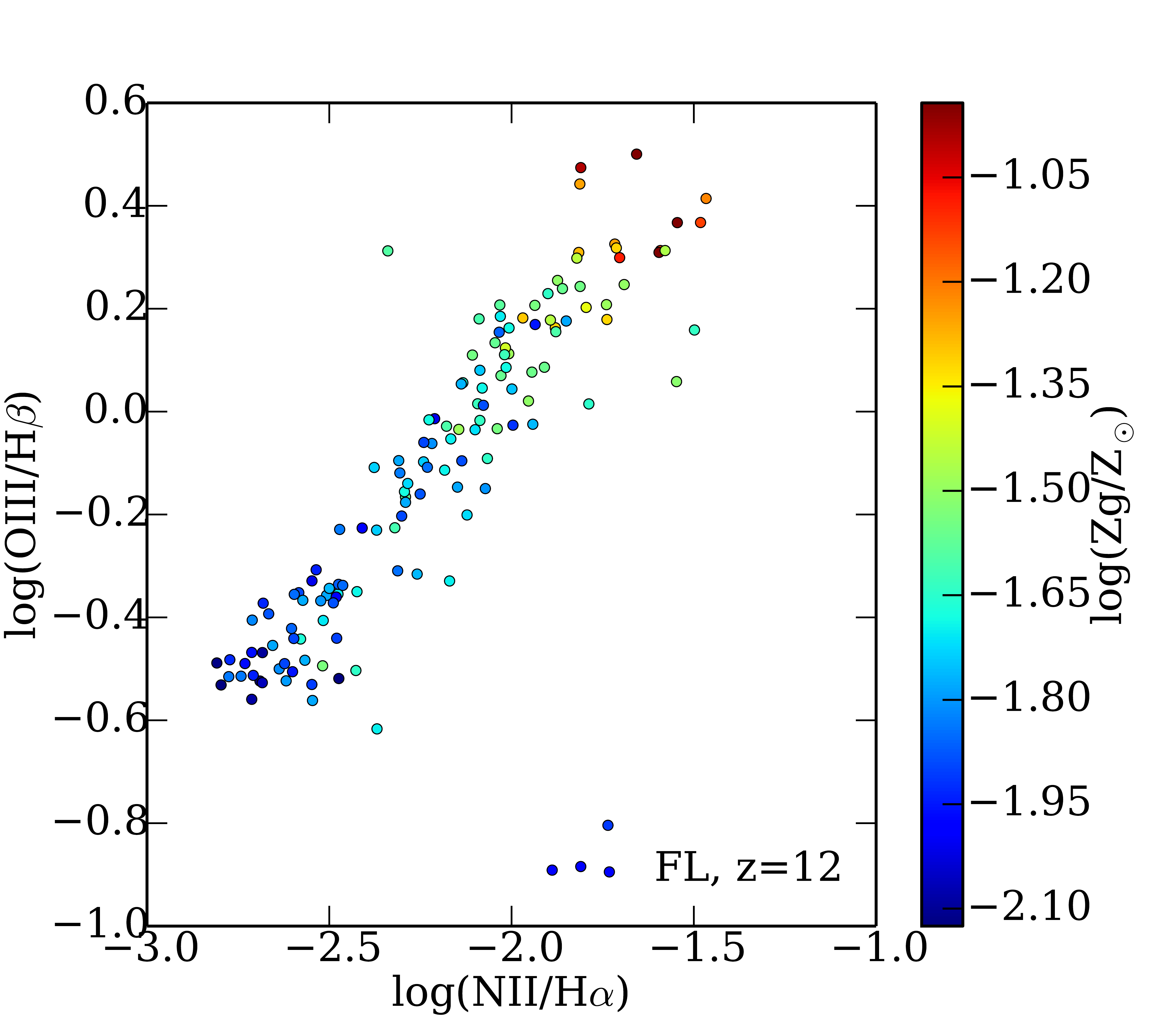}	
	 \caption{BPT diagrams at $z=8$ and 12 coloured by the mean gas metallicity (Z$_{\rm g}$)} 
	  \label{fig:A4}
\end{figure}

\begin{figure}
	\includegraphics[width=\columnwidth]{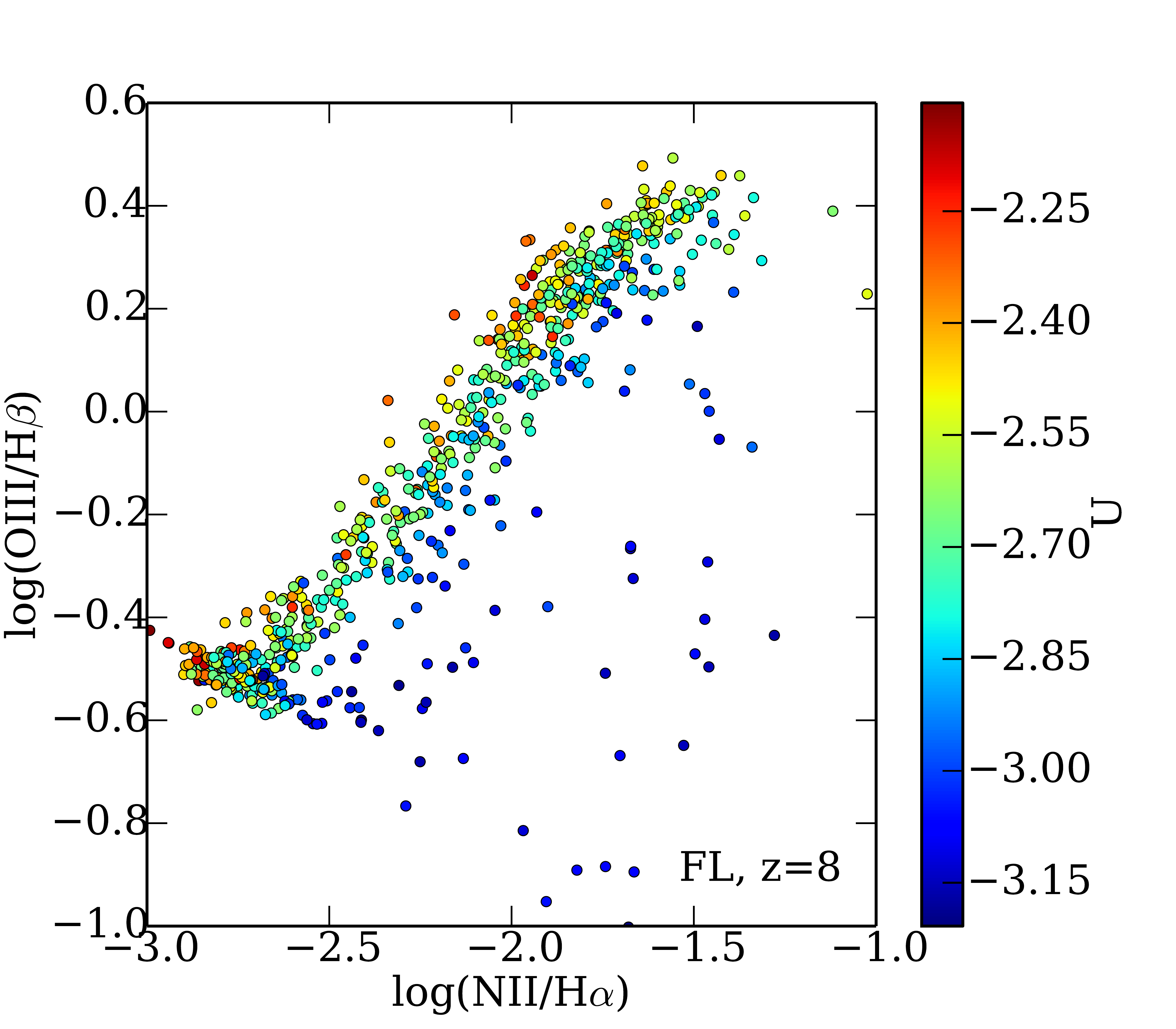}	
	\includegraphics[width=\columnwidth]{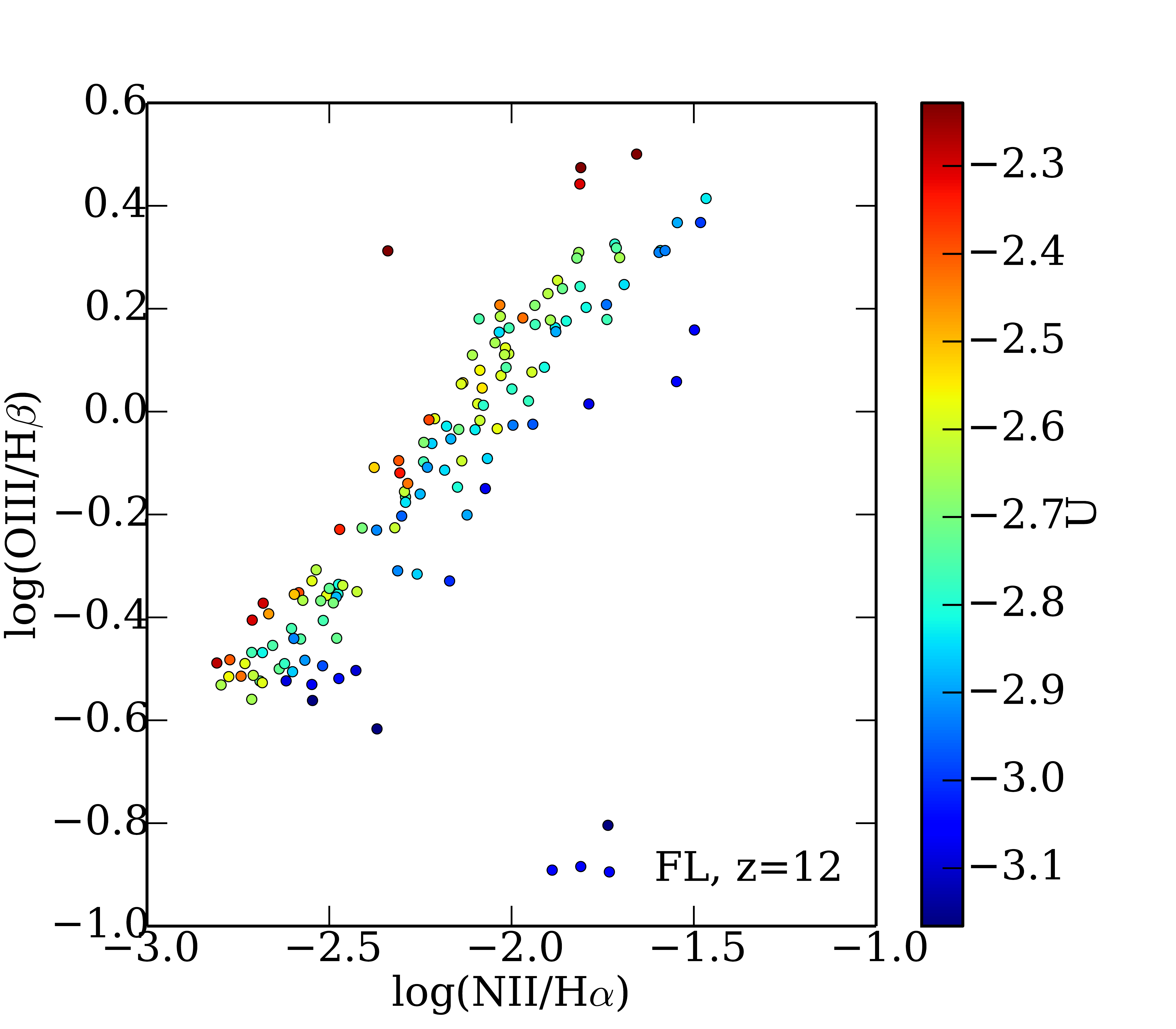}	
	 \caption{BPT diagrams at $z=8$ and 12 coloured by the mean ionization parameter (U).}
	  \label{fig:A5}
\end{figure}

\Fig{A2} illustrates the lack of evolution of the equivalent width of the combined H$\alpha$+NII lines (\Fig{EWHaNII}). 
\Fig{A3} shows the evolution of the scaling relation between the luminosity in the V band and the stellar mass of the galaxy. It is similar to the evolution of the $\Ms$-M$_{1500}$ relation shown in \Fig{MsMUVSSFR}.
\Fig{A4} and \Fig{A5} show the BPT diagrams (\Fig{BPT}) for $z=8$ and $z=12$. There is no clear evolution.

\bsp	
\label{lastpage}
\end{document}
